\documentclass[11pt, oneside]{article}   	
\usepackage{slashed}
\usepackage{jheppub}
\usepackage{amsmath}
\usepackage{amssymb}
\usepackage{algorithm}
\usepackage{algpseudocode}
\usepackage{amsfonts}
\usepackage{amsthm}
\usepackage{amsbsy}
\usepackage{array}
\usepackage{mathtools}
\usepackage[usenames,dvipsnames]{xcolor}
\usepackage{subcaption}
\usepackage{dsdshorthand}
\usepackage{graphicx}
\usepackage[vcentermath]{youngtab}
\usepackage{tikz}
\usetikzlibrary{snakes}
\usetikzlibrary{decorations}
\usetikzlibrary{trees}
\usetikzlibrary{decorations.pathmorphing}
\usetikzlibrary{decorations.markings}
\usetikzlibrary{external}
\usetikzlibrary{intersections}
\usetikzlibrary{shapes,arrows}
\usetikzlibrary{arrows.meta}
\usetikzlibrary{calc}
\usetikzlibrary{shapes.misc}
\usetikzlibrary{decorations.text}
\usetikzlibrary{backgrounds}
\usetikzlibrary{fadings}
\usetikzlibrary{tikzmark,calc,arrows,shapes,decorations.pathreplacing}
\tikzset{
        cross/.style={cross out, draw=black, minimum size=2*(#1-\pgflinewidth), inner sep=0pt, outer sep=0pt},
	branchCut/.style={postaction={decorate},
		snake=zigzag,
		decoration = {snake=zigzag,segment length = 2mm, amplitude = 2mm}	
    }}

\newcommand{\bea}{\setlength\arraycolsep{2pt} \begin{eqnarray}}
\newcommand{\eea}{\end{eqnarray}}
\def\ft#1#2{{\textstyle{\frac{\scriptstyle #1}{\scriptstyle #2} } }}
\def\fft#1#2{{\frac{#1}{#2}}}

\newcommand{\baa}{\begin{align}}
\newcommand{\eaa}{\end{align}}

\def\lsim{\mathrel{\hbox{\rlap{\lower.55ex \hbox{$\sim$}} \kern-.3em \raise.4ex \hbox{$<$}}}}
\def\gsim{\mathrel{\hbox{\rlap{\lower.55ex \hbox{$\sim$}} \kern-.3em \raise.4ex \hbox{$>$}}}}

\makeatletter
\def\@fpheader{\ }
\makeatother

\title{Effective field theory bootstrap, large-N $\chi$PT and holographic QCD}
\author{Yue-Zhou Li$^{1,2}$}
\affiliation{${}^1$Department of Physics, Princeton University, Princeton, NJ 08544, USA\\
${}^2$Department of Physics, McGill University, 3600 Rue University, Montr\'eal, H3A 2T8, QC Canada
}
\emailAdd{liyuezhou@princeton.edu}
\date{}
\abstract{We review the effective field theory (EFT) bootstrap by formulating it as an infinite-dimensional semidefinite program (SDP), built from the crossing symmetric sum rules and the S-matrix primal ansatz. We apply the program to study the large-$N$ chiral perturbation theory ($\chi$PT) and observe excellent convergence of EFT bounds between the dual (rule-out) and primal (rule-in) methods. This convergence aligns with the predictions of duality theory in SDP, enabling us to analyze the bound states and resonances in the ultra-violet (UV) spectrum. Furthermore, we incorporate the upper bound of unitarity to uniformly constrain the EFT space from the UV scale $M$ using the primal method, thereby confirming the consistency of the large-$N$ expansion. In the end, we translate the large-$N$ $\chi$PT bounds to constrain the higher derivative corrections of holographic QCD models.}

\begin{document}

\maketitle
\pagenumbering{roman}
\setcounter{page}{2}
\newpage
\pagenumbering{arabic}
\setcounter{page}{1}

\section{Introduction}

For long distances beyond a certain characteristic scale $1/M$, low-energy effective field theories (EFTs) are utilized to describe physical processes and predict observables. In these scenarios, ultra-violet (UV) effects manifest as tails shaped by higher-dimensional operators, which are suppressed by $1/M$. However, at short distances, UV effects intensify and become non-negligible, raising a profound question: What is the allowable space of EFTs that ensure a consistent UV completion, such as in quantum gravity?
Particularly regarding completion to quantum gravity, there are numerous intriguing conjectures and arguments, known as the swampland program \cite{Ooguri:2006in,Brennan:2017rbf,Palti:2019pca,vanBeest:2021lhn}. Examples include the weak gravity conjecture \cite{ArkaniHamed:2006dz,Harlow:2022ich,Cheung:2018cwt}, the distance conjecture \cite{Ooguri:2006in,Grimm:2018ohb}, and others. These are inspired by string theory and studies of black hole physics, and they provide conceptual criteria for gaining insights into this question.

Even without the gravitational degree of freedom, this question remains profound and warrants further investigation. It is known that EFTs without gravity can still be pathological. Having oversized Wilson coefficients \cite{Camanho:2014apa} or possessing the wrong sign for some EFT Wilson coefficients \cite{Adams:2006sv} can violate causality.

The EFT bootstrap program has recently been developed to quantitatively and systematically explore this question, assuming the unitarity and causality of the underlying UV theory above $M$, as well as Regge boundedness \cite{deRham:2017avq,Caron-Huot:2020cmc,Tolley:2020gtv,Arkani-Hamed:2020blm,Bellazzini:2020cot,Chiang:2021ziz}. The strategy involves studying $2$-to-$2$ scattering amplitudes in EFTs, denoted as $\mathcal{M}_{\rm EFT}$, and then searching for the allowed space of Wilson coefficients. Causality and Regge boundedness provide a bridge between the EFT amplitudes $\mathcal{M}_{\rm EFT}$ and the underlying UV amplitudes, which is known as the dispersive sum rules. Built upon dispersive sum rules, the unitarity can then be used to optimally carve out the EFT space. This whole procedure is known as the dual bootstrap algorithm, as it rigorously rules out disallowed values of Wilson coefficients. The dual EFT bootstrap can also incorporate dynamical gravity \cite{Caron-Huot:2021rmr,Caron-Huot:2021enk,Caron-Huot:2022ugt,Caron-Huot:2022jli}, thereby providing sharp bounds on some of the swampland conjectures \cite{Henriksson:2022oeu,Hong:2023zgm}. There are many relevant works utilizing this idea to constrain EFTs and their UV completions, see, e.g., \cite{Sinha:2020win,Alberte:2020bdz,Li:2021lpe,Bern:2021ppb,Haldar:2021rri,Raman:2021pkf,Henriksson:2021ymi,Zahed:2021fkp,Alberte:2019xfh, AccettulliHuber:2020oou,Alberte:2021dnj,Chowdhury:2021ynh,Bellazzini:2021oaj,Zhang:2021eeo,Wang:2020jxr, Trott:2020ebl,deRham:2017zjm,deRham:2018qqo,Wang:2020xlt,deRham:2022gfe,Noumi:2022wwf,Albert:2022oes,Fernandez:2022kzi,Bellazzini:2023nqj,Albert:2023jtd,Chen:2023bhu}.

However, there is a weakness in the current version of the dual EFT bootstrap. Typically, to optimize the EFT bounds, it is essential to ensure that we measure only a finite number of Wilson coefficients in which we are interested. Meanwhile, the null constraints should be employed \cite{Caron-Huot:2020cmc,Tolley:2020gtv,Arkani-Hamed:2020blm,Bellazzini:2020cot,Chiang:2021ziz}. These null constraints are constructed using crossing symmetry, a crucial ingredient of quantum causality. They complement the original dispersive sum rules because the latter are not fully crossing symmetric and thus lack important information. These requirements are usually met using the improved sum rules \cite{Caron-Huot:2021rmr}, which subtract the forward limit expansions from the original sum rules. This subtraction ensures they measure only those Wilson coefficients that saturate specific Regge boundedness. However, this procedure can be vulnerable to loop effects since the forward limit scale might compete with the loop expansions. This competition essentially hinders an efficient generalization that powerfully constrains EFTs at the loop level (for exploration on this subject, see, e.g., \cite{Bellazzini:2021oaj,Li:2022aby}). Even at tree-level, this forward limit subtraction can sometimes complicate numerical exploration.

In this paper, we will explore the numerical usage of the crossing symmetric dispersive sum rules \cite{Auberson:1972prg,Mahoux:1974ej,Sinha:2020win}, which automatically incorporate the crossing symmetry. In other words, the null constraints are encoded in the crossing symmetric sum rules, and these sum rules only retain a finite number of Wilson coefficients. We will show that this type of sum rule is an optimized version of the ``improved sum rules", as it fulfills all the requirements that the ``improved sum rules" satisfy and it is free of any forward limit subtractions. Although our discussions in this paper are limited to tree-level, we believe that the crossing symmetric sum rules are an excellent playground for understanding the loop effects of EFT bounds.

There is a different approach termed the primal S-matrix bootstrap \cite{Paulos:2017fhb}. The primal bootstrap is a powerful tool to constrain quantum field theories (QFT) non-perturbatively: it is built upon an appropriate ansatz of the S-matrix designed to obey causality and directly searches for optimal couplings under unitarity constraints. The term ``primal" indicates that this method is used to rule in allowed values of couplings. Although the S-matrix bootstrap was proposed to constrain the dynamics of non-perturbative QFTs (with a large amount of applications, see, .e.g., \cite{Cordova:2018uop,Guerrieri:2018uew,EliasMiro:2019kyf,Cordova:2019lot,Bercini:2019vme,Guerrieri:2021ivu,Guerrieri:2022sod,Karateev:2022jdb,Marucha:2023vrn,acanfora2023bounds,Acanfora:2023axz}), it can be easily adapted for studying EFTs, e.g., \cite{Guerrieri:2020bto,Chen:2022nym,Haring:2022sdp}. Some natural questions then arise: how do the resulting EFT bounds from the primal method compare to those from the dual method? In what sense are the primal and dual bootstraps dual to each other in the context of optimization theory? Regarding the first question, intuitively, we expect that when both methods are applied to the same EFT with the same assumptions, their resulting bounds should converge to each other, eventually leading to the optimal constraints of EFTs. This convergence has indeed been observed in relevant investigations for scalar EFTs \cite{Chen:2022nym}. For the second question, it has been recognized that the primal bootstrap is the ``primal problem" in optimization theory such as semidefinite program (SDP) \cite{Guerrieri:2020kcs,Guerrieri:2021tak}, and therefore one can construct its optimization Lagrangian and identify the ``dual problem". This ``dual problem" should be the method one can construct using the dispersive sum rules as the dual EFT bootstrap \cite{EliasMiro:2022xaa}.

In this paper, we will show that the EFT bootstrap can indeed be formulated as an infinite-dimensional SDP problem. This guarantees that the dual and primal bounds should converge to each other, as long as the strong duality is satisfied \cite{wolkowicz2012handbook}. The strong duality condition can be translated to conditions of EFT bootstrap, which then indicates that we can extract UV physics from the primal solutions under the guidance of the dual extremal functionals.

We then apply the EFT bootstrap with the crossing symmetric sum rules, now formulated as an SDP, to the case study of large-$N$ chiral perturbation theory ($\chi$PT). $\chi$PT is an EFT that describes light meson physics, arising from chiral symmetry breaking in the low-energy regime of quantum chromodynamics (QCD). Large-$N$ $\chi$PT is the meson EFT that emerges from large-$N$ QCD \cite{Coleman:1980mx}, which generalizes the colour group to SU$(N)$ with $N\rightarrow\infty$ \cite{t1993planar}, and provides a qualitative understanding of many aspects of hadron physics. The dual EFT bootstrap program for large-$N$ $\chi$PT was initiated in \cite{Albert:2022oes}, and it is still under active investigation \cite{Fernandez:2022kzi,Albert:2023jtd,Ma:2023vgc}. One advantage of studying the large-$N$ limit is that it retains only tree-level physics, where the dual EFT bootstrap is efficient. For finite $N$, such as in real QCD, $\chi$PT may become strongly coupled around the threshold and therefore the loop cannot be neglected. To our knowledge, in this case, only primal studies has been be performed \cite{Guerrieri:2020bto,He:2023lyy}. On the other hand, the large-$N$ limit is also expected to have a string description \cite{t1993planar}, so it might provide rich ``experiments" for understanding quantum gravity and holography \cite{Maldacena:1997re,Gubser:1998bc,Witten:1998qj}. In this paper, we will study large-$N$ $\chi$PT using both dual and primal methods, and we observe excellent convergence. We also extract the physical spectrum from our primal solutions and the dual functionals, not only confirming some understandings from \cite{Albert:2022oes,Fernandez:2022kzi}, but also revealing some novel and hidden physics that seems to be accessible only when using the primal method.

Typically, for $\mathcal{O}(p^4)$ Wilson coefficients, a segment of the boundary corresponds precisely to the Skyrme model \cite{Albert:2022oes}. However, as the Wilson coefficients increase beyond the kink \cite{Albert:2022oes,Fernandez:2022kzi}, the Skyrme model is excluded. On the other hand, large-$N$ QCD is known to have string and holographic descriptions, such as the well-known Witten-Sakai-Sugimoto model \cite{Witten:1998zw,Sakai:2004cn,Sakai:2005yt}, which yields precisely the Skyrme model at low energy \cite{Sakai:2004cn,Sakai:2005yt}. It is intriguing to translate the constraints on large-$N$ $\chi$PT to see if there are any problems with holographic QCD models. Not surprisingly, all known holographic QCD models produce the Skyrme model that exists below the kink, and is thus consistent. However, these models have all been analyzed only at the leading order as EFTs of gauge fields. We will show that including the higher dimensional operators on the gravity side leads to a large-$N$ $\chi$PT that deviates from the Skyrme model, controlled by the bulk Wilson coefficients. This allows us to translate the large-$N$ $\chi$PT bounds to constrain the bulk EFTs of gauge fields on non-trivial backgrounds.

The rest of the paper is summarized as follows. In section \ref{sec: EFT bootstrap}, we review the basic ideas of dispersive sum rules and provide the construction of crossing symmetric sum rules. After reviewing the structures of amplitudes and the unitarity constraints, we focus on the positive unitarity condition, explaining why the EFT bootstrap is an infinite-dimensional SDP and how we can construct its Lagrangian formulation and derive physics from SDP duality. In section \ref{sec: review chPT}, we review large-$N$ $\chi$PT, including its Lagrangian, the flavour structure of the pion amplitudes, and the partial waves and unitarity conditions. We then explicitly construct the associated crossing symmetric dispersive sum rules and the primal S-matrix ansatz as our dual problem setup and the primal problem setup, respectively. In section \ref{sec: bounds}, we obtain the EFT bounds using both the dual and primal algorithms and present the convergence between the two methods. We also display the spectral density and S-matrix, which are numerically solved using the primal methods for saturating certain EFT bounds. Using a simple sample bound, we demonstrate that modifying the Regge behaviour of the primal ansatz does not alter the resulting bounds, as long as it stays below the Regge boundedness assumption. As an ad hoc approach, we incorporate the upper bound of unitarity to uniformly bound the $\mathcal{O}(p^4)$ Wilson coefficients in terms of the cut-off scale $M$, which does not contradict the large-$N$ bound, thereby confirming the consistency of the large-$N$ limit. In section \ref{sec: holographic QCD}, we study holographic QCD. We include the higher dimensional operators built from gauge fields and show that holographic QCD with higher derivative terms produces the most general $\chi$PT Lagrangian at order $\mathcal{O}(p^4)$. We verify that the Witten-Sakai-Sugimoto model has no issues with the leading string corrections. Afterwards, we translate the chiral EFT bounds to constrain the $5D$ EFT with double gauge fields. We conclude the paper in section \ref{sec: summary}. Appendix \ref{app: EFT} formulates other EFT bootstrap scenarios as SDP; appendix \ref{app: projector} records the SU$(N_f)$ projectors that we used to organize the pion amplitudes and partial waves; and appendix \ref{app: fixing-parameter} provides more details on the bootstrap Lagrangian for fixing one parameter and bounding another.

\section{EFT bootstrap and SDP problem}
\label{sec: EFT bootstrap}
\subsection{Dispersive sum rules}

\subsubsection{Basic ideas}

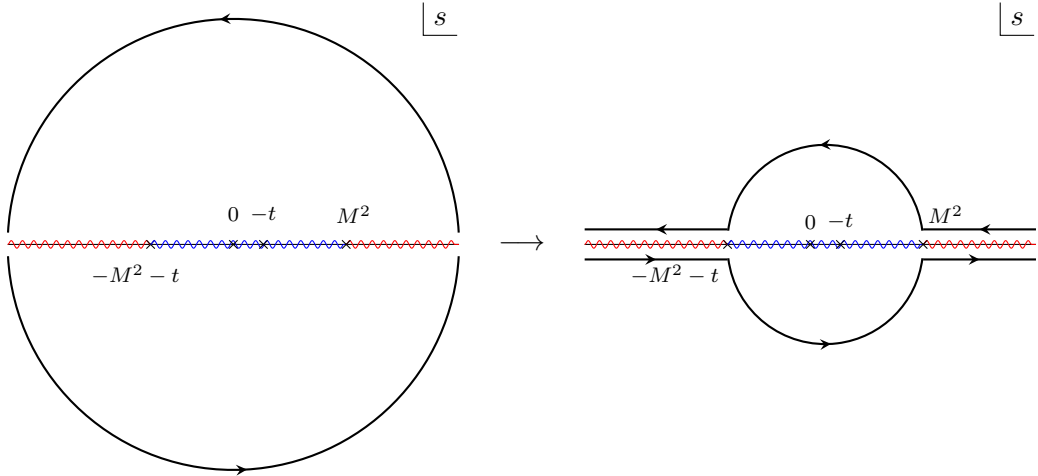
\begin{figure}[t]
\centering 
\begin{tikzpicture}[decoration={markings, 
    mark= at position 0.52 with {\arrow{stealth}}}]
	\draw (-3,0) -- (3,0);
    \draw[thick, rotate=3, postaction={decorate}] (-3,0) arc (-180:-6:3) ;
    \draw[thick, rotate=3, postaction={decorate}] (3,0) arc (0:174:3) ;
    \draw[red, decorate, decoration={snake=zigzag,segment length=1.5mm, amplitude=0.5mm}]       (-3,0) -- (-1.1,0);
    \fill (-1.1,0) node[red, cross=2pt] {};
    \draw[blue, decorate, decoration={snake=zigzag,segment length=1.5mm, amplitude=0.5mm}]       (-1.1,0) -- (1.5,0);
	\draw[red, decorate, decoration={snake=zigzag,segment length=1.5mm, amplitude=0.5mm}]       (1.5,0) -- (3,0);
    \fill (0,0) node[cross=2pt] {};
	\draw (0.4,0) node[yshift=0.4cm] {\scriptsize $-t$};
    \fill (0.4,0) node[cross=2pt] {};
    \fill (1.5,0) node[cross=2pt] {};
	\node (a) at (2.75,3) {$s$};
	\draw (a.north west) -- (a.south west) -- (a.south east);
	\draw (0,0) node[yshift=0.4cm] {\scriptsize $0$};
	\draw (1.5,0) node[xshift=0.1cm,yshift=0.4cm] {\scriptsize $M^2$};
	\draw (-1.5,0) node[yshift=-0.4cm,xshift=0.2cm] {\scriptsize $-M^2 -t$};
	\end{tikzpicture}
    \raisebox{86pt}{$\quad\longrightarrow\quad$}
\begin{tikzpicture}
	\draw (-3,0) -- (3,0);
    \draw[white, thick, rotate=-3, postaction={decorate}] (3,0) arc (0:-174:3) ;
    \draw[white, thick, rotate=3, postaction={decorate}] (3,0) arc (0:174:3) ;
    \draw[thick, rotate=7, postaction={decorate}, decoration={markings, 
        mark= at position 0.52 with {\arrow{stealth}}}] (-1.105,-0.05) arc (-180:-14:1.3) ;
    \draw[thick, rotate=7, postaction={decorate}, decoration={markings, 
        mark= at position 0.52 with {\arrow{stealth}}}] (1.5,0) arc (0:166:1.3) ;
	\draw[thick, thick, postaction={decorate}, decoration={markings, 
            mark= at position 0.5 with {\arrow{stealth}}}] (1.5,-0.2) --(3,-0.2) ;
	\draw[thick, thick, postaction={decorate}, decoration={markings, 
            mark= at position 0.5 with {\arrow{stealth}}}] (-3,-0.2) --(-1.1,-0.2) ;
    \draw[thick, postaction={decorate}, decoration={markings, 
            mark= at position 0.5 with {\arrow{stealth}}}] (3,0.2) --(1.5,0.2) ;
	\draw[thick, thick, postaction={decorate}, decoration={markings, 
            mark= at position 0.5 with {\arrow{stealth}}}] (-1.1,0.2) --(-3,0.2) ;
    \draw[red, decorate, decoration={snake=zigzag,segment length=1.5mm, amplitude=0.5mm}]       (-3,0) -- (-1.1,0);
    \fill (-1.1,0) node[cross=2pt] {};
	\draw[red, decorate, decoration={snake=zigzag,segment length=1.5mm, amplitude=0.5mm}]       (1.5,0) -- (3,0);
	 \draw[blue, decorate, decoration={snake=zigzag,segment length=1.5mm, amplitude=0.5mm}]       (-1.1,0) -- (1.5,0);
    \fill (0,0) node[cross=2pt] {};
	\draw (0.4,0) node[yshift=0.3cm] {\scriptsize $-t$};
    \fill (0.4,0) node[cross=2pt] {};
    \fill (1.5,0) node[cross=2pt] {};
	\draw (1.5,0) node[xshift=0.3cm,yshift=0.4cm] {\scriptsize $M^2$};
	\node (a) at (2.75,3) {$s$};
	\draw (a.north west) -- (a.south west) -- (a.south east);
	\draw (0,0) node[yshift=0.3cm] {\scriptsize $0$};
	\draw (-1.5,0) node[yshift=-0.4cm,xshift=-0.3cm] {\scriptsize $-M^2-t$};
	\end{tikzpicture}
\caption{The contour deformation leads to the sum rules given by eq.~\eqref{eq: dispersion}. The red branch cut represents the UV branch cut, which is beyond our knowledge, while the blue branch cut represents the low-energy cut contributed by loop effects in low-energy EFTs. The final contour relates low-energy EFT data along the arcs to the discontinuity along the UV branch cuts.}\label{fig: contour s-u}
\end{figure}

One essential component of EFT bootstrap is the dispersive sum rules. The strategy is to design vanishing integral identities along a large circle at infinity in the complex $s$ plane, e.g., 
\be
B_k(p^2)= \oint_{\infty} \fft{ds}{4\pi i} \fft{\mathcal{M}(s,-p^2)}{s^{k+1}}\equiv 0\,,\quad k\geq k_0 \in \mathbb{Z}\,,\label{eq: integral identity}
\ee
where $k_0$ is the ceiling of the Regge spin $J_0$ for UV amplitudes
\be
\Big|\mathcal{M}(s,t)\Big|_{|s|\rightarrow\infty} \sim |s|^{J_0} < |s|^{k_0}\,,\quad \text{for fixed}\, t<0\,.\label{eq: usual Regge}
\ee
Causality is also assumed, which is thought to imply both the crossing symmetry and the analyticity of the S-matrix in the complex $s$ plane, except for poles and branch cuts in the real axis (see \cite{Correia:2020xtr,Mizera:2023tfe} for more details on the analyticity of the S-matrix). Analyticity allows us to deform the contour of integral identities \eqref{eq: integral identity} towards the real axis, with a smaller arc within the regime where low-energy EFTs remain valid ($|s|< M^2$)\footnote{We consider EFTs in which the mass of the particles, $m$, is much smaller than $M$. Therefore, we treat the scattering in EFTs as massless scattering, and ignore all the anomalous thresholds.}, as illustrated in Fig \ref{fig: contour s-u}.

This procedure establishes the dispersion relations (or dispersive sum rules) that relate low-energy to high-energy physics
\be
-B_k(p^2)\Big|_{\text{low arc}} = B_k(p^2)\Big|_{\rm high}\,.\label{eq: dispersion}
\ee
This gives
\be
\oint_{|s|<M^2}\fft{ds}{4\pi i}\fft{\mathcal{M}(s,-p^2)}{s^{k+1}}=\int_{M^2}^{\infty}\fft{ds}{2\pi} \fft{{\rm Disc} \mathcal{M}(s,-p^2)}{s^{k+1}} + \text{$u$-channel}\,,\label{eq: sum rules usual}
\ee
where the discontinuity is defined by
\be
{\rm Disc} f = \fft{f(s+i 0)-f(s-i 0)}{2i}\,.
\ee
On the other hand, crossing symmetry allows for the relationship between the $u$-channel contribution and the $s$-channel contribution. Crossing symmetry enables us to modify and improve the dispersive sum rules into a more convenient and powerful basis by subtracting the null constraints \cite{Caron-Huot:2020cmc}. For instance, it is demonstrated in \cite{Caron-Huot:2021rmr} that improved spin-$k$ sum rules can be constructed by subtracting the forward-limit expansions of higher spin sum rules (i.e., $k^\prime>k$). This method leaves only a finite number of Wilson coefficients with Regge spin $k$ in the low-energy measurement, which is exceptionally beneficial\footnote{The Regge spin of a Wilson coefficient can be defined by the exponent of $s$ in the fixed-$t$ Regge limit of the associated tree-level amplitude in EFTs. For example, for a higher-dimensional operator giving amplitudes $gs^k$ in the fixed-$t$ Regge limit, we say the Regge spin of $g$ is $k$.}. For example, the improved spin-$2$ sum rule for four-dimensional scalar EFT is given by \cite{Caron-Huot:2021rmr}
\be
B_2^{\rm imp}=\oint_{\infty} \fft{ds}{4\pi i} \Big(\fft{(2s-p^2)}{s^2(s-p^2)^{2}}\mathcal{M}(s,-p^2)-\fft{(4s-3p^2)p^4}{s^4(s-p^2)^2}\mathcal{M}(s,0) -\fft{2p^6}{s^3 (s^2-p^4)}\partial_{p^2} \mathcal{M}(s,0)\Big)\equiv 0\,,
\ee
which at low-energy (tree-level) only measures gravity and Wilson coefficients of dimension-$8$ and dimension-$10$ operators\footnote{Those operators contribute to low-energy amplitudes by $(s^2+t^2+u^2)g^{{\rm dim}8} + s t u \, g^{{\rm dim}10}\subset \mathcal{M}$.}
\be
B_2^{\rm imp}\big|_{\rm low}=\fft{8\pi G}{p^2}+2 g^{{\rm dim}8} + p^2 g^{{\rm dim}10}\,. \label{eq; scalar example}
\ee

However, it is obvious that the construction of this improvement is heavily dependent on details; see \cite{Caron-Huot:2022ugt,Caron-Huot:2022jli} for more complicated scattering processes. In addition, the construction relies on the forward-limit expansion, making the loop effects vulnerable. In this note, we will instead use the crossing symmetric sum rules \cite{Auberson:1972prg,Mahoux:1974ej,Sinha:2020win}, which we will introduce momentarily, to build causality (i.e., analyticity plus crossing symmetry) directly into the dispersive sum rules.

\subsubsection{Crossing symmetric sum rules}

We review the crossing symmetric sum rules in this section\footnote{We are grateful to Simon Caron-Huot for drawing our attention to this fantastic construction.}. The essence lies in the analytic and crossing-symmetric parameterization of the Mandelstam variables in terms of a complex variable $z$ and an auxiliary momentum $p$
\be
s(z,p)=-\fft{3p^2 z}{1+z+z^2}\,,\quad t(z,p)=s(z\, \xi,p)\,,\quad u(z,p)=s(z \,\xi^2,p)\,,
\ee
where $\xi=e^{2/3 i \pi}$ and $0<p^2\leq M^2/3$. In terms of the complex $z$-plane, the Mandelstam variables are geometrically symmetrical, with an angular difference of $2/3\pi$ from each other. The Regge limit in this parametrization is associated with special points on the unit circle. For example, the fixed-$t$ Regge limit $|s|\rightarrow\infty$ corresponds to $z=\xi^2$; other channels follow similarly. To build the crossing symmetric sum rules, it is necessary to study the full crossing symmetric amplitudes and find the crossing symmetric kernel that can perform the sufficient subtractions in the Regge limit. This kernel is easy to construct, from which we obtain the fixed-$p$ identity
\be
B_k(p^2)=\oint_{z=1,\xi,\xi^2} \fft{dz}{4\pi i}\, \mathcal{K}_k(z) \mathcal{M}^{\rm sym}(z,p^2)\equiv 0\,,\quad k\geq k_0\,\,\text{and}\,\, k\in2\mathbb{Z}\,,\label{eq: cross-sym identity}
\ee
where
\be
\mathcal{K}_k(z)= (-1)^{\fft{k}{2}}3^{1-\frac{3 k}{2}} \left(z^3+1\right) p^{-2 k} z^{-\frac{3 k}{2}-1} \left(1-z^3\right)^{k-1}\,.
\ee
Note that the integration contour consists of small circles surrounding the Regge limit point. The transformation of the integration variable to $s$ yields
\be
B_k(p^2)=\oint \fft{ds}{4\pi i}\,  s^{-\frac{3 k}{2}-1} \left(3 p^2+2 s\right) \left(p^2+s\right)^{\frac{k}{2}-1} \mathcal{M}^{\rm sym}(s,p^2)\,.
\ee
Here, the Mandelstam variables $t$ and $u$ are parameterized in terms of $s$ by
\be
t=-\fft{s(p^2+s-\sqrt{s-3p^2}\sqrt{s+p^2})}{2(s+p^2)}\,,\quad u=-\fft{s(p^2+s+\sqrt{s-3p^2}\sqrt{s+p^2})}{2(s+p^2)}\,.
\ee

In order to deform the contour in \eqref{eq: cross-sym identity} and obtain the sum rule, we must understand the analyticity of $\mathcal{M}^{\rm sym}(z,p^2)$ in terms of the complex variable $z$. We have three pieces of UV branch cuts: $s\geq M^2, u\geq M^2$ for fixed-$t$; $s\geq M^2, t\geq M^2$ for fixed-$u$; and $t\geq M^2, u\geq M^2$ for fixed-$s$. In the complex $z$ plane, these branch cuts all reside on the unit circle and sandwich the Regge limit points in the associated channels. The UV branch cuts are summarized as follows

\begin{itemize}
\item UV branch cuts
\item[1.] Fixed-$u$
\be
& s\geq M^2:\quad 2\pi -\cos^{-1}(-\ft{1}{2}(1+3p^2))\leq  {\rm Arg}\,z< \fft{4\pi}{3}\,,\nn\\
& t \geq M^2:\quad \fft{4\pi}{3}< {\rm Arg}\,z \leq \fft{2\pi}{3} +\cos^{-1}(-\ft{1}{2}(1+3p^2))\,,\nn\\
& \text{Regge point}:\quad {\rm Arg}\,z=\fft{4\pi}{3}\,.\nn
\ee
 
\item[2.] Fixed-$t$
\be
& s\geq M^2:\quad \fft{2\pi}{3}< {\rm Arg}\,z\leq \cos^{-1}(-\ft{1}{2}(1+3p^2))\,,\nn\\
& u \geq M^2:\quad \fft{4\pi}{3}-\cos^{-1}(-\ft{1}{2}(1+3p^2))\leq {\rm Arg}\,z < \fft{2\pi}{3}\,,\nn\\
& \text{Regge point}:\quad {\rm Arg}\,z=\fft{2\pi}{3}\,.\nn
\ee

\item[3.] Fixed-$s$
\be
& t\geq M^2:\quad \fft{2\pi}{3} -\cos^{-1}(-\ft{1}{2}(1+3p^2))\leq  {\rm Arg}\,z< 0\,,\nn\\
& u \geq M^2:\quad 0< {\rm Arg}\,z \leq  -\fft{2\pi}{3} +\cos^{-1}(-\ft{1}{2}(1+3p^2))\,,\nn\\
& \text{Regge point}:\quad {\rm Arg}\,z=0\,.\nn
\ee
\end{itemize} 

\begin{figure}[t]
\centering 
\begin{tikzpicture}
  \draw[dashed] (0,0) circle (2cm);

\foreach \angle/\label in {0/\scriptsize fixed-$s$,120/\scriptsize fixed-$u$,240/\scriptsize fixed-$t$} {
  \fill (\angle:2cm) circle (1pt); 
  
  \draw[postaction={decorate, decoration={
            markings,
            mark=at position 0.5 with {\arrow{stealth}}
          }}] (\angle:2cm) circle (0.15);
  \node at (\angle:2.6cm) {\label};
}

 \fill (0,0) node[cross=2pt] {};
 \draw (0,0) node[yshift=0.3cm] {\scriptsize $0$};
  \fill (-2.6,1.5) node[cross=2pt] {};
 \draw (-2.6,1.5) node[yshift=0.3cm] {\scriptsize $\infty$};
 \draw[red, decorate, decoration={snake=zigzag,segment length=1.5mm, amplitude=0.5mm}] (-145:2cm) arc (-145:-124:2cm);
  \draw[red, decorate, decoration={snake=zigzag,segment length=1.5mm, amplitude=0.5mm}] (-116:2cm) arc (-116:-95:2cm);
    \draw[red, decorate, decoration={snake=zigzag,segment length=1.5mm, amplitude=0.5mm}] (25:2cm) arc (25:4:2cm);
     \draw[red, decorate, decoration={snake=zigzag,segment length=1.5mm, amplitude=0.5mm}] (-4:2cm) arc (-4:-25:2cm);
      \draw[red, decorate, decoration={snake=zigzag,segment length=1.5mm, amplitude=0.5mm}] (116:2cm) arc (116:95:2cm);
        \draw[red, decorate, decoration={snake=zigzag,segment length=1.5mm, amplitude=0.5mm}] (124:2cm) arc (124:145:2cm);
        \draw[blue, decorate, decoration={snake=zigzag,segment length=1.5mm, amplitude=0.5mm}] (-95:2cm) arc (-95:-25:2cm);
        \draw[blue, decorate, decoration={snake=zigzag,segment length=1.5mm, amplitude=0.5mm}] (145:2cm) arc  (145:215:2cm);
\draw[blue, decorate, decoration={snake=zigzag,segment length=1.5mm, amplitude=0.5mm}] (95:2cm) arc (95:25:2cm);
            \draw[blue, decorate, decoration={snake=zigzag,segment length=1.5mm, amplitude=0.5mm}]       (-2.5,0) -- (0,0);
            \draw[blue, decorate, decoration={snake=zigzag,segment length=1.5mm, amplitude=0.5mm}]       (1.25,2.16506) -- (0,0);
             \draw[blue, decorate, decoration={snake=zigzag,segment length=1.5mm, amplitude=0.5mm}]       (1.25,-2.16506) -- (0,0);
             \node (a) at (2.25,2) {$z$};
             \draw (a.north west) -- (a.south west) -- (a.south east);
	\end{tikzpicture}
\raisebox{68pt}{$\quad\longrightarrow\quad$}
\begin{tikzpicture}
  \draw[dashed] (0,0) circle (2cm);
\foreach \angle/\label in {0/\scriptsize fixed-$s$,120/\scriptsize fixed-$u$,240/\scriptsize fixed-$t$} {
  \fill (\angle:2cm) circle (1pt); 

  \node at (\angle:2.6cm) {\label};
}

 \fill (0,0) node[cross=2pt] {};
 \draw (0,0) node[yshift=0.3cm] {\scriptsize $0$};
  \fill (-2.6,1.5) node[cross=2pt] {};
 \draw (-2.6,1.5) node[yshift=0.3cm] {\scriptsize $\infty$};
 \draw[red, decorate, decoration={snake=zigzag,segment length=1.5mm, amplitude=0.5mm}] (-145:2cm) arc (-145:-124:2cm);
  \draw[red, decorate, decoration={snake=zigzag,segment length=1.5mm, amplitude=0.5mm}] (-116:2cm) arc (-116:-95:2cm);
 \draw[thick,postaction={decorate}, decoration={markings, 
            mark= at position 0.5 with {\arrow{stealth}}}] (-145:1.9cm) arc (-145:-124:1.9cm);
            \draw[thick,postaction={decorate}, decoration={markings, 
            mark= at position 0.5 with {\arrow{stealth}}}] (-124:2.1cm) arc (-124:-145:2.1cm);
   \draw[thick, postaction={decorate}, decoration={markings, 
            mark= at position 0.5 with {\arrow{stealth}}}] (-116:1.9cm) arc (-116:-95:1.9cm);
            \draw[thick, postaction={decorate}, decoration={markings, 
            mark= at position 0.5 with {\arrow{stealth}}}] (-95:2.1cm) arc (-95:-116:2.1cm);
    \draw[thick,postaction={decorate}, decoration={markings, 
            mark= at position 0.5 with {\arrow{stealth}}}] (-0.82,-0.57)--(-1.56,-1.09);
              \draw[thick,postaction={decorate}, decoration={markings, 
            mark= at position 0.5 with {\arrow{stealth}}}] (-0.17,-1.89) -- (-0.09,-1);
              \draw[thick, postaction={decorate}, decoration={markings, 
            mark= at position 0.5 with {\arrow{stealth}}}] (-95:1cm) arc (-95:-145:1cm);
              \draw[thick,postaction={decorate}, decoration={markings, 
            mark= at position 0.5 with {\arrow{stealth}}}] (-1.72,-1.2)--(-2.05,-1.43);
              \draw[thick,postaction={decorate}, decoration={markings, 
            mark= at position 0.5 with {\arrow{stealth}}}]  (-0.22,-2.5)--(-0.18,-2.09) ;
    \draw[red, decorate, decoration={snake=zigzag,segment length=1.5mm, amplitude=0.5mm}] (25:2cm) arc (25:4:2cm);
     \draw[red, decorate, decoration={snake=zigzag,segment length=1.5mm, amplitude=0.5mm}] (-4:2cm) arc (-4:-25:2cm);
      \draw[thick,postaction={decorate}, decoration={markings, 
            mark= at position 0.5 with {\arrow{stealth}}}] (4:1.9cm) arc (4:25:1.9cm);
            \draw[thick,postaction={decorate}, decoration={markings, 
            mark= at position 0.5 with {\arrow{stealth}}}] (25:2.1cm) arc (25:4:2.1cm);
   \draw[thick, postaction={decorate}, decoration={markings, 
            mark= at position 0.5 with {\arrow{stealth}}}] (-25:1.9cm) arc (-25:-4:1.9cm);
            \draw[thick, postaction={decorate}, decoration={markings, 
            mark= at position 0.5 with {\arrow{stealth}}}] (-4:2.1cm) arc (-4:-25:2.1cm);
    \draw[thick,postaction={decorate}, decoration={markings, 
            mark= at position 0.5 with {\arrow{stealth}}}] (0.91,-0.42)--(1.72,-0.8);
              \draw[thick,postaction={decorate}, decoration={markings, 
            mark= at position 0.5 with {\arrow{stealth}}}] (1.9,-0.89) -- (2.27,-1.06);
              \draw[thick, postaction={decorate}, decoration={markings, 
            mark= at position 0.5 with {\arrow{stealth}}}] (25:1cm) arc (25:-25:1cm);
              \draw[thick,postaction={decorate}, decoration={markings, 
            mark= at position 0.5 with {\arrow{stealth}}}] (1.72,0.8)--(0.91,0.42);
              \draw[thick,postaction={decorate}, decoration={markings, 
            mark= at position 0.5 with {\arrow{stealth}}}]  (2.27,1.06) --(1.9,0.89);
      \draw[red, decorate, decoration={snake=zigzag,segment length=1.5mm, amplitude=0.5mm}] (116:2cm) arc (116:95:2cm);
        \draw[red, decorate, decoration={snake=zigzag,segment length=1.5mm, amplitude=0.5mm}] (124:2cm) arc (124:145:2cm);
         \draw[thick,postaction={decorate}, decoration={markings, 
            mark= at position 0.5 with {\arrow{stealth}}}] (124:1.9cm) arc (124:145:1.9cm);
            \draw[thick,postaction={decorate}, decoration={markings, 
            mark= at position 0.5 with {\arrow{stealth}}}] (145:2.1cm) arc (145:124:2.1cm);
   \draw[thick, postaction={decorate}, decoration={markings, 
            mark= at position 0.5 with {\arrow{stealth}}}] (95:1.9cm) arc (95:116:1.9cm);
            \draw[thick, postaction={decorate}, decoration={markings, 
            mark= at position 0.5 with {\arrow{stealth}}}] (116:2.1cm) arc (116:95:2.1cm);
    \draw[thick,postaction={decorate}, decoration={markings, 
            mark= at position 0.5 with {\arrow{stealth}}}] (-1.56,1.09)--(-0.82,0.57);
              \draw[thick,postaction={decorate}, decoration={markings, 
            mark= at position 0.5 with {\arrow{stealth}}}]  (-0.09,1)--(-0.17,1.89) ;
              \draw[thick, postaction={decorate}, decoration={markings, 
            mark= at position 0.5 with {\arrow{stealth}}}] (145:1cm) arc (145:95:1cm);
              \draw[thick,postaction={decorate}, decoration={markings, 
            mark= at position 0.5 with {\arrow{stealth}}}](-2.05,1.43)-- (-1.72,1.2);
              \draw[thick,postaction={decorate}, decoration={markings, 
            mark= at position 0.5 with {\arrow{stealth}}}] (-0.18,2.09)-- (-0.22,2.5);
            \draw[blue, decorate, decoration={snake=zigzag,segment length=1.5mm, amplitude=0.5mm}]       (-2.5,0) -- (0,0);
            \draw[blue, decorate, decoration={snake=zigzag,segment length=1.5mm, amplitude=0.5mm}]       (1.25,2.16506) -- (0,0);
             \draw[blue, decorate, decoration={snake=zigzag,segment length=1.5mm, amplitude=0.5mm}]       (1.25,-2.16506) -- (0,0);
             \node (a) at (2.25,2) {$z$};
             \draw (a.north west) -- (a.south west) -- (a.south east);
              \draw[red, decorate, decoration={snake=zigzag,segment length=1.5mm, amplitude=0.5mm}] (124:2cm) arc (124:145:2cm);
         \draw[blue, decorate, decoration={snake=zigzag,segment length=1.5mm, amplitude=0.5mm}] (-95:2cm) arc (-95:-25:2cm);
        \draw[blue, decorate, decoration={snake=zigzag,segment length=1.5mm, amplitude=0.5mm}] (145:2cm) arc  (145:215:2cm);
\draw[blue, decorate, decoration={snake=zigzag,segment length=1.5mm, amplitude=0.5mm}] (95:2cm) arc (95:25:2cm);
\end{tikzpicture}
\caption{The analytic structures and the contour deformation for crossing symmetric sum rules. The red branch cut represents the UV branch cut, and the blue branch cut represents the low-energy cut contributed by loop effects in low-energy EFTs.}\label{fig: contour cross}
\end{figure}
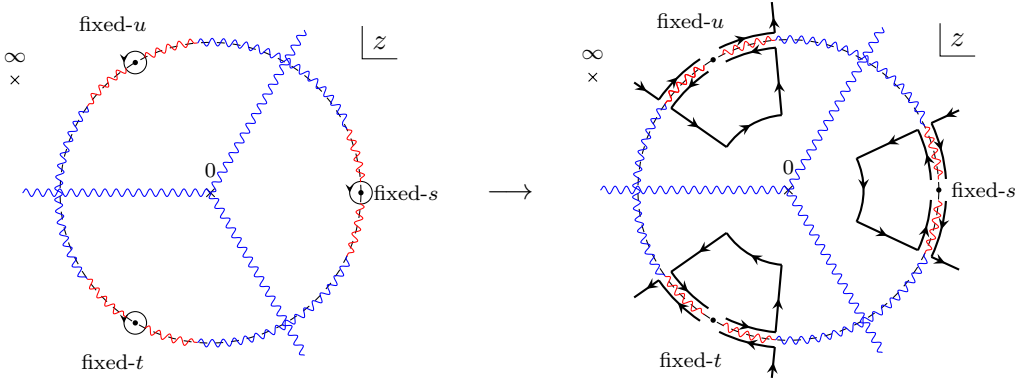

In this paper, we consider only the tree-level at low-energy. However, it is instructive to analyze the low-energy analyticity when loops are present. A salient feature of the crossing symmetric representation is its clear distinction between the UV branch cuts and the low-energy branch cuts, which are not even connected for, e.g., $s<3p^2$. The low-energy branch cuts, such as $0\leq s<3p^2$, extend from $|z|=0$ to $|z|=\infty$ at three specific angles. For $3p^2<s<M^2$, the branch cut resides on the unit circle complement to the UV branch cuts \footnote{We are grateful to Cyuan-Han Chang for pointing out the error in the previous version of \eqref{eq: low-energy cut} and Fig \ref{fig: contour cross}. In the previous version, the author forgot to complete the low-energy branch cuts.}
\begin{itemize}
\item Low-energy branch cuts
\be
& 0<s< 3p^2: \quad |z|\in (0,\infty)\,,{\rm Arg}\,z = \pi\,,\nn\\
& 0<t<3p^2: \quad |z|\in (0,\infty)\,,{\rm Arg}\,z = -\fft{\pi}{3}\,,\nn\\
& 0<u< 3p^2: \quad |z|\in (0,\infty)\,,{\rm Arg}\,z = \fft{\pi}{3}\,\nn\\
&3p^2<s<M^2:\quad |z|=1\,,\cos^{-1}(-\ft{1}{2}(1+3p^2))<{\rm Arg} z< 2\pi -\cos^{-1}(-\ft{1}{2}(1+3p^2))\,,\nn\\
&3p^2<t<M^2:\quad |z|=1\,,\fft{2\pi}{3} +\cos^{-1}(-\ft{1}{2}(1+3p^2))<{\rm Arg} z< \fft{2\pi}{3} -\cos^{-1}(-\ft{1}{2}(1+3p^2))\,,\nn\\
&3p^2<u<M^2:\quad |z|=1\,,-\fft{2\pi}{3}+\cos^{-1}(-\ft{1}{2}(1+3p^2))<{\rm Arg} z< \fft{4\pi}{3} -\cos^{-1}(-\ft{1}{2}(1+3p^2))\,.\label{eq: low-energy cut}
\ee
\end{itemize}
It is important to note that at tree-level, the low-energy massless poles $s=t=u=0$ are all located on $|z|=0$ and $|z|=\infty$.

We are now ready to deform the contour and build the sum rules, as shown in Fig. \ref{fig: contour cross}.  In this figure, the UV contour takes the discontinuities along the red UV branch cut, while the low-energy contour is stretched both inwards and outwards from the unit circle.

We continue to have \eqref{eq: dispersion}, and more specifically, it now yields
\be
& \oint_{C_{\rm IR}}\fft{dz}{4\pi i}\, \mathcal{K}_k(z)\mathcal{M}^{\rm sym}(z,p^2)=\int_{M^2}^{\infty}\fft{ds}{2\pi}s^{-\frac{3 k}{2}-1} \left(3 p^2+2 s\right) \left(p^2+s\right)^{\frac{k}{2}-1} {\rm Disc} \mathcal{M}^{\rm sym}(s,p^2)\,.\label{eq: cross-sym sum rule}
\ee
It is worth noting that we express the UV part in terms of $s$, which will be convenient when performing the partial-wave expansion. The sum rule \eqref{eq: cross-sym sum rule} is constructed to be crossing symmetric, and it naturally subtracts all null constraints in a nonlinear manner. Indeed, for example, the low-energy part for scalar EFT at tree-level is precisely the same as in \eqref{eq; scalar example}, as observed by \cite{deRham:2022gfe}, but we are not taking any forward-limit! We argue that the crossing symmetric sum rule is a more natural and well-defined approach for addressing low-energy loops.

\subsubsection{Manipulate sum rules using functionals}

To harness the powerful capabilities of dispersive sum rules, it is instructive to build functionals that manipulate sum rules and measure the interesting couplings at low-energy
\be
-\sum_k\mathcal{F}_k\circ B(p^2)\Big|_{\text{low arc}} =\sum_k \mathcal{F}_k\circ B_k(p^2)\Big|_{\rm high}\,.\label{eq: functional}
\ee
To derive constraints on EFTs, one can the search for functionals that optimize quantities at low-energy, subject to the unitarity that we will introduce later. Generally, we can define the functionals by smearing the sum rules against wave functions
\be
\mathcal{F}\circ f :=  \int_0^{p_{\rm max}^2} dp^2 \psi(p^2)f\,,
\ee
where $p_{\rm max}^2=M^2$ for \eqref{eq: sum rules usual}, while $p_{\rm max}^2=M^2/3$ for \eqref{eq: cross-sym sum rule}. 

There are two kinds of functionals in the literature, which we call the forward-limit functional and the impact parameter functional \cite{Caron-Huot:2021rmr}. The forward-limit function is achieved by taking the wave function $\psi(p^2)= \sum_i c_i \partial_{p^2}^i$, which then performs the forward-limit expansion; on the other hand, the impact parameter functional measures the sum rules at small impact parameter $b\sim 1/M$.
\begin{itemize}
\item Forward-limit functional
\be
\psi(p^2)=\sum_i c_i \partial_{p^2}^i \rightarrow \mathcal{F}_{p^2\rightarrow0}\circ f:=\sum_{i} c_i\, \partial_{p^2}^i f\,.
\ee

\item Impact parameter functional

$\psi(p^2)$ has finite support in the momentum space and decays fast enough in the impact parameter space $\psi(b):=\int d^{d-2}p\, e^{ib\cdot p} \psi(p^2)$; such functionals are usually chosen as\footnote{It is worth noting that one has to pay attention when choosing the starting point of the polynomial $p^{i_0}$, which controls the numerics in the large impact parameter regime $b\rightarrow\infty$ \cite{Caron-Huot:2021rmr}.}
\be
\psi(p^2)= \sum_i c_i\, p^i\,,\quad i\in \mathbb{Z}\,,
\ee
as well as its variants for numerical benefits \cite{Caron-Huot:2021rmr,Caron-Huot:2022ugt,Caron-Huot:2022jli}.
\end{itemize}

The forward-limit functional is much simpler and requires fewer computational resources, therefore it is more often used in EFTs without graviton. However, this functional can be singular at low-energy when dealing with graviton propagation due to the $1/t$ graviton pole at low-energy. In contrast, the impact parameter functional would suppress the graviton pole, making the gravitational low-energy behaviour well-defined under its action\footnote{In $4D$, the graviton pole is not completely resolved. Nevertheless, the divergence can be improved from polynomial divergence to the logarithmic IR divergence $\log M/m_{\rm IR}$ using the functionals integrated from $m_{\rm IR}^2$ \cite{Caron-Huot:2021rmr,Caron-Huot:2022ugt}. This logarithmic IR divergence reflects the behaviour of the classical Newton potential. As a consequence, the functional becomes ineffective beyond $b_{\rm max}\sim 1/m_{\rm IR}$, a region that we should simply discard.}. In addition, the impact parameter functional also provides a bonus for allowing one to weaken the Regge boundedness assumption \eqref{eq: usual Regge}. The essential reason behind this bonus is that smearing amplitudes against the fast decay wave function $\psi(b)$ would suppress the higher spin contributions at high energy, effectively enhancing the Regge behaviour under the smearing \cite{Caron-Huot:2022ugt} (see also \cite{Haring:2022cyf} for a more evident proof).

In this paper, we do not include the graviton, therefore we will use the forward-limit functional for quick convergence of numerics.

\subsection{Low-energy amplitudes}

It is worth noting that at this stage, the dispersive sum rules formally utilize the full amplitudes, even along the low-energy arc. For low-energy part, since this arc is inside the EFT regime, we may expect to replace the amplitudes there with EFT amplitudes. The simplest situation is that the underlying theory is weakly coupled at low energy, where we can replace low-energy amplitudes along the small arc with tree-level EFT amplitudes. These amplitudes contain only simple poles, allowing us to evaluate the arc integral by picking up the residues of simple poles. This is the simplest case that has been extensively studied. Generally, we have
\be
\mathcal{M}_{\rm EFT}(s,t;\mu) + \mathcal{M}_{\rm match}(M^2,\mu) = \mathcal{M}(s,t)\,,
\ee
where the full amplitudes $\mathcal{M}(s,t)$ are expanded in a Taylor series in terms of $1/M$. The EFT amplitudes are computed using the effective Lagrangian, which is dependent on the scale through logarithmic structures such as $\log(s/\mu)$ and $\log(m^2/\mu)$. An additional matching piece often appears because the order of Taylor expansions and the integrals do not commute, i.e.,
\be
\big(\int d^d x \mathcal{L}\big)\Big|_{\rm expansion} \neq \int d^d x \mathcal{L}_{\rm eff}\,.
\ee
The matching piece contains terms like $\log(M^2/\mu)$. For simplicity, we often choose $\mu=M^2$, which gives us a simpler relation
\be
\mathcal{M}_{\rm EFT}(s,t;\mu=M^2)=\mathcal{M}(s,t)\,.
\ee
Therefore, the dispersive sum rules measure the Wilson coefficients at scale $\mu=M^2$; it is then necessary to apply the renormalization group equation to evolve Wilson coefficients back to other scales.

In this paper, we will study the large-$N$ chiral EFT, therefore the tree-level approximation is sufficient.

\subsection{Unitarity constraints}

How do we use \eqref{eq: dispersion} to constrain EFTs in terms of the UV amplitudes when the details of the UV theory are absent? It is instructive to study amplitudes at high energy using the partial wave expansion
\be
\mathcal{M}(s,t)=s^{\fft{4-d}{2}}\sum_\rho \fft{2^{d+1}(2\pi)^{d-1}{\rm dim}\rho}{{\rm Vol}S^{d-1}} a_\rho(s) \,\pi_\rho\big(1+\fft{2t}{s}\big)\,,
\ee
where $\rho$ labels the irreducible representation of ${\rm SO}(d)$ and $\pi_\rho$ is the associated partial waves \cite{Caron-Huot:2022jli}, and we slip off the indices of possible global symmetry\footnote{See \cite{Buric:2023ykg} for excellent constructions of partial waves for arbitrary spin using the representation theory.}. The unitarity then implies a strong constraint on the partial wave coefficients
\be
|1+i a_\rho(s)|^2\leq 1\,.\label{eq: full unitary}
\ee

This nonlinear unitarity condition implies the positivity constraint
\be
{\rm Disc}\, a_\rho(s) \geq 0\,.
\ee
This is the scenario that gives rise to the positivity bounds considered in the literature \cite{deRham:2017avq}. If the couplings of the underlying theory are weak enough that the quadratic terms $|a_\rho(s)|^2 \ll {\rm Disc}\,a_\rho(s)$ can be ignored, then the positivity condition robustly constrains the low-energy EFTs. However, it is important to understand that a weakly coupled EFT does not necessarily mean its UV completion will also be weakly coupled. The essential condition is the existence of a parametrically small but positive parameter $0<g\ll 1$ in low-energy EFTs that can be measured by dispersive sum rules. This leads us to
\be
0<g = \sum_{\rho}\int_{M^2}^{\infty} dm^2 Y_\rho(m^2) {\rm Disc}\,a_\rho(m^2)\ll 1\,,
\ee
where $Y_\rho(m^2)$ is any appropriate function which is generated by functionals acting on partial waves. This obviously shows that ${\rm Disc}\,a_\rho$ has to be parametrically small as $g$. In this case, the optimal bounds on Wilson coefficients will be scaling with $g$. Gravitational EFTs studied in \cite{Caron-Huot:2022ugt,Caron-Huot:2022jli} (and their couplings to scalar and photon \cite{Henriksson:2022oeu,Hong:2023zgm,Hamada:2023cyt}) fall into this category, where the parametrically small parameter is the Newton constant $G_N=1/M_{\rm pl}^{d-2}\ll 1/M^{d-2}\ll 1$; Large-$N$ chiral EFT that will be studied in the following sections also falls into this category, where the small parameter is the inverse of the pion decay constant $1/f_\pi^2\sim 1/N\ll 1$ \cite{Coleman:1980mx} (see section \ref{sec: review chPT} for more details).

In other cases, the full unitarity \eqref{eq: full unitary} will be providing more stringent constraints on low-energy EFTs, such as scalar EFTs studied in \cite{Chen:2022nym}. It turns out that the full unitarity \eqref{eq: full unitary} can be linearized by formalizing it as a positive matrix \cite{Paulos:2017fhb}
\be
|1+i a_\rho(s)|^2\leq 1 \rightarrow 
\left(
\begin{array}{cc}
 {\rm Disc}\, a_\rho(s) & {\rm Re}\, a_\rho(s) \\
{\rm Re}\, a_\rho(s) & 2-{\rm Disc}\, a_\rho(s) \\
\end{array}
\right)\succeq 0 \,,
\ee
where ${\rm Re}\,a_\rho(s)=1/2 \big(a_\rho(s+i 0)+a_\rho(s-i 0)\big)$. It is then easy to see that for small ${\rm Disc}\, a_\rho(s)$ we only need to consider the first diagonal element of this matrix; on the other hand, if ${\rm Disc}\,a_\rho(s)$ is not necessary small but ${\rm Re}\,a_\rho(s)$ is small, we can also ignore the off-diagonal terms and impose the linear constraint $0\leq {\rm Disc}\,a_\rho(s)\leq 2$ (which is considered in, e.g., \cite{Chen:2022nym,Chiang:2022ltp,Chiang:2022jep,Chen:2023bhu}).

These discussions allow us to classify the scenarios of EFT bootstrap, which invoke different unitarity conditions according to the basic assumptions (we follow the terminology invented in \cite{Chen:2022nym})

\begin{itemize}

\item[I.] Positivity

\be
{\rm Disc} \,a_\rho (s) \geq 0\,,\quad \exists\,\, g\in \mathcal{P}_{\rm EFT}\,\,\, \text{where}\,\,\, 0<g\ll 1\,\, \text{and}\,\, g\subset B_k(p^2)\Big|_{\text{low arc}}\,.\nn
\ee

\item[II.] Linear unitarity

\be
0\leq {\rm Disc}\,a_\rho(s)\leq 2\,,\quad \text{if}\,\,\big({\rm Re}\,a_\rho(s)\big)^2 \ll {\rm Disc}\,a_\rho(s)\,.\nn
\ee

\item[III.] Nonlinear unitarity

\be
\left(
\begin{array}{cc}
 {\rm Disc}\, a_\rho(s) & {\rm Re}\, a_\rho(s) \\
{\rm Re}\, a_\rho(s) & 2-{\rm Disc}\, a_\rho(s) \\
\end{array}
\right)\succeq 0\,.\nn
\ee

\end{itemize} 

From now on, for simplicity, we will denote the discontinuity as the imaginary part, using the notation ${\rm Disc}\rightarrow{\rm Im}$, when there is no confusion.

\subsection{EFT bootstrap as infinite dimensional SDP}

\subsubsection{Semidefinite programming}

Since the unitarity of the S-matrix can be expressed as a semidefinite matrix, carving out the allowed space of EFTs can then be transformed into the semi-definite program (SDP), subject to those unitarity constraints. In this section, we provide a crash course on SDP. We will then show that EFT bootstrap is an infinite-dimensional SDP.

The SDP can be formulated as the following primal optimization procedure \cite{wolkowicz2012handbook,Simmons-Duffin:2015qma}

\begin{itemize}
\item Primal problem
\be
\text{Minimize}\,\quad &c\cdot x\,\,\, \text{over}\,\,\, x\in\mathbb{R}^N\,,\nn\\
\text{Subject to}\,\quad &X: = \sum_{i=1}^N A_i x_i - C \succeq 0\,,\,\,\,X\in \mathcal{S}^K\nn\\
&  B^T x=b\,,\,\,\, b\in \mathbb{R}^P\,, B\in \mathbb{R}^{P\times N} \nn
\ee
\end{itemize}
In this algorithm, $\mathcal{S}^K$ is the space of $K\times K$ symmetric real matrices. 

The primal S-matrix bootstrap \cite{Paulos:2017fhb}, as applied to EFTs, falls into this problem with infinite dimensions. For simplicity, we only consider the positivity constraint. We can approximate the full amplitude (which is valid for both UV and low-energy EFT) by using an infinite number of analytic but simpler functions $\mathcal{M}_i$ with the assumed analyticity and Regge behaviour
\be
\mathcal{M}(s,t) = \sum_{i=1} x_i \, \mathcal{M}_i(s,t)\,.
\ee
The positivity of the unitary condition now becomes
\be
\sum_{i=1} x_i \,{\rm Im}\,a_\rho^i(s)\geq 0\,,\quad \text{for all allowed}\, J\geq 0 \in \rho\,\, \text{and for all}\, s\geq M^2\,\text{with fixed $t<0$}\,,
\ee
where $a_\rho^i$ is the partial wave coefficients contributed by $\mathcal{M}^i$. If we imagine that we put ${\rm Im}\,a_\rho(s)$ at all values of $s\geq M^2$ for all spins into an infinite-dimensional diagonal matrix, the primal EFT bootstrap is, in principle, an infinite-dimensional primal problem with $K, N=\infty$ when taking $C\equiv 0$.\footnote{We can also formally think about $C$ as $|\mathcal{M}||^2/2$, which is infinitesimally small in the positivity scenario.} Here, $B^T x =b$ is simply the normalization condition for fixing a particular Wilson coefficient, and we can minimize the target Wilson coefficient by choosing appropriate $c$. This is because any Wilson coefficient can be represented in terms of a linear combination of $x_i$ by taking the low-energy limit of the full amplitudes. However, in practice, we cannot reach $N, K=\infty$ in numerics. Instead, one takes a maximal value of $N_{\rm max}$, and we also consider up to a certain $J_{\rm max}$, imposing unitarity for a finite but large number of $s$-grids \cite{Paulos:2017fhb}. This truncation procedure gives a well-defined optimization, after which one must extrapolate the bound \cite{Chen:2022nym,Guerrieri:2021ivu,Guerrieri:2022sod}.

Duality plays an important role in SDP. Typically, the primal problem has a dual formulation, known as the dual problem
\begin{itemize}
\item Dual problem
\be
\text{Maximize}\,\quad &{\rm Tr}(CY)+b\cdot y\,\,\, \text{over}\,\,\, y\in\mathbb{R}^P\,\text{and}\, Y\in \mathcal{S}^K\,,\nn\\
\text{Subject to}\,\quad &{\rm Tr} (A_i Y) + \sum_{j=1}^P B_{ij}y_j = c_i\,,\,\,\, \text{and}\,\,\, Y  \succeq 0\,.\nn
\ee
\end{itemize}
We can easily construct the dual version for primal EFT bootstrap with positivity that we described previously. The maximization target is $b\cdot y$ since we choose $C\equiv 0$, we then have
\be
\int_{M^2}^{\infty} ds\sum_{\rho} {\rm Im}\, a_\rho^i(s) Y_\rho(s) + \sum_j B_{ij}y_j=c_i\,,\quad Y_\rho(s)\geq 0\,.\label{eq: dual}
\ee
where we have already explicitly evaluated the trace by integrating over $s\geq M^2$ and summing over all spins in the irreducible representation. Note that in this language, we use $Y_\rho(s)$ to denote an infinite-dimensional matrix in which the diagonal elements take the values of all $s\geq M^2$ and spins in $\rho$. What does \eqref{eq: dual} mean? It becomes clear if we dot \eqref{eq: dual} into $x$, we find
\be
\int_{M^2}^\infty ds \sum_{\rho} {\rm Im}\, a_\rho(s) Y_\rho(s) = c\cdot x- b\cdot y\,.\label{eq: dual rep}
\ee
In other words, the dual problem is to find a positive function $Y_\rho(s)$ such that its average against the spectral density can represent low-energy Wilson coefficients $c\cdot x$, and we bound $c\cdot x$ by maximizing $b\cdot y$ according to the positivity. The dual problem \eqref{eq: dual rep} can obviously be achieved by using the dispersive sum rules, which is precisely the dual bootstrap algorithm studied in \cite{Caron-Huot:2020cmc}. In this algorithm, the function $Y_\rho(s)$ can be constructed by acting the functionals on the UV part of sum rules
\be
\sum_k \mathcal{F}_k\circ B_k(p^2)\Big|_{\rm high}= \int_{M^2}^\infty ds \sum_{\rho} {\rm Im}\, a_\rho(s) Y_\rho(s)\,.
\ee
Other scenarios can also be formulated as SDPs, see \cite{Guerrieri:2021tak,EliasMiro:2022xaa} and appendix \ref{app: EFT} for further discussions.

It is also natural to ask whether the primal and dual problems yield the same optimal value for our target. This question can be answered by the duality theory in SDP, which we will review in the following subsections.

\subsubsection{The Lagrangian formulation, SDP duality and physical implications}

SDP has the Lagrangian formulation, which manifests the logics of optimization. A SDP is described by the following Lagrangian
\be
\mathcal{L}(x,y,Y)&=c\cdot x-{\rm Tr}(X Y)+ (b-B^T x)\cdot y\nn\\
& =({\rm Tr}( CY)+ b\cdot y \big)-\sum_{i=1}^N \big({\rm Tr}(A_i Y) + \sum_{j=1}^P B_{ij}y_j-c_i\big)x_i\,,\label{eq: Lag SDP}
\ee
where the first line is intended to manifest the primal problem, while the second line targets the dual problem. Equality is achieved using the expression  $X=\sum_{i=1}^N A_i x_i-C$ along with some basic algebra. It's worth noting that we don't use the subject identity, because we want to emphasize that every component of SDP can be seen from the Lagrangian. Using this Lagrangian, the primal problem can be formulated as \cite{wolkowicz2012handbook}
\be
\mathcal{P}=\min_{x} \Big(\sup_{Y \succeq 0} \mathcal{L}(x,y,Y)\Big)\,,\quad \text{for}\,\, X \succeq0\,,
\ee
where we have used 
\be
\sup_{Y\succeq 0}\big(-{\rm Tr}(X Y)\big) = 
\left\{
\begin{array}{ll}
0\,, & \text{if } X\succeq 0\,,\\
\infty\,, & \text{otherwise}\,.
\end{array}
\right.\qquad
 \sup_{Y\succeq 0} \big( (b-B^T x)\cdot y\big) =
\left\{
\begin{array}{ll}
0\,, & \text{if } b=B^T x\,,\\
\infty\,, & \text{otherwise}\,.
\end{array}
\right.
\ee
The dual problem can then be constructed by interchanging the ordering of ``minimize'' and ``maximize'', namely
\be
\mathcal{D}=\max_{Y\succeq 0}\Big(\inf_{x} \mathcal{L}(x,y,Y)\Big)\,.
\ee
This indeed gives rise to the standard dual problem by noting
\be
\inf_{x}\mathcal{L}(x,y,Y)= 
\left\{
\begin{array}{ll}
{\rm Tr}(CY)+b\cdot y\,, & \text{if }\,\, {\rm Tr}(A_i Y)+\sum_{j=1}^P B_{ij}y_j-c_i =0\,,\\
\infty\,, & \text{otherwise}\,.
\end{array}
\right.
\ee

According to our previous discussions, we can then immediately traslate \eqref{eq: Lag SDP} to the Lagrangian of positivity EFT bootstrap, built from any reasonable functionals acting on the dispersive sum rules
\be
\mathcal{L}_{I}= \lambda_{\mathcal{F}} -\mathcal{F}\circ B(p^2)\,.
\ee
The first term is the dual objective, and the second term represents functionals acting on sum rules for our targets, giving rise to \eqref{eq: functional}. This Lagrangian is compact and is the guide throughout this paper. To be more concrete, let's say we want to minimize a Wilson coefficient $g_{\mathcal{F}}$ in terms of $g_0>0$. We can simply expand $\mathcal{F}\circ B(p^2)$
\be
\mathcal{F}\circ B(p^2)= - g_{\mathcal{F}} + \lambda_{\mathcal{F}}\, g_0 + \int_{M^2}^{\infty} ds \sum_\rho {\rm Im}\,a_\rho(s)Y_\rho(s)\,.
\ee
We then have
\be
\mathcal{L}_I = g_{\mathcal{F}} + \lambda_{\mathcal{F}}(1-g_0) - \int_{M^2}^{\infty}ds \sum_\rho {\rm Im}\,a_\rho(s)Y_\rho(s)\,.
\ee
We can now easily read off either the primal or dual algorithm. The primal problem is straightforward to read; we fix $g_0=1$ and then minimize $g_{\mathcal{F}}$ subject to the positivity ${\rm Im}\, a_\rho(s)\geq 0$. It is worth noting that setting $g_0=1$ is not physical; this is a scaling trick to deal with the numerics, and one can always set other values of $g_0$. The physical result is the lower bound of $g_{\mathcal{F}}/g_0$; similarly, the solved ${\rm Im}\,a_\rho$ is also in the unit of $g_{10}$. For the dual problem, we maximize $\lambda_{\mathcal{F}}$ subject to $\mathcal{F}\circ B(p^2)\equiv 0$ with $Y_\rho(s)\succeq 0$. Therefore, we have
\be
 g_{\mathcal{F}}-\lambda_{\mathcal{F}}g_0=\int_{M^2}^{\infty} ds \sum_{\rho}{\rm Im}\,a_\rho(s) Y_\rho(s) \geq 0 \,\,\, \rightarrow \,\,\, g_{\mathcal{F}}\geq (\max \lambda_{\mathcal{F}}) g_0\,.
\ee
This is precisely the dual algorithm proposed by \cite{Caron-Huot:2020cmc}.

A natural question arises: For the same bootstrap problem, do the primal and dual methods yield the same constraints? This question can be answered using the duality theory of SDP. In general, if we find feasible solutions for both the primal and dual problems, we have weak duality, which states that the primal bound is always greater than the dual bound, leading to a duality gap as follows
\be
c\cdot x - \big({\rm Tr}(C Y)-b\cdot y\big) = {\rm Tr}(XY) \geq 0\,.
\ee
This is the reason the primal bound is always referred to as the rule-in bound, while the dual bound is termed as the rule-out bound. To ensure the duality gap vanishes, we clearly need $XY\equiv 0$ and the Slater's condition $X\succ 0$ or $Y\succ 0$ \cite{wolkowicz2012handbook}. This implies that a solution to the problem must satisfy
\be
X\succ 0\,,\quad Y\equiv 0\,,\qquad \text{or}\qquad Y\succ 0\,,\quad X\equiv 0\,.
\ee
Physically, this condition is powerful. For example, numerical conformal bootstrap employs $Y\equiv 0$ to identify the physical spectrum, a method known as the extremal functional method \cite{Poland:2010wg,ElShowk:2012hu}. In terms of positivity EFT bootstrap, the second possible condition is trivial, it simply provides a free field theory solution. The nontrivial physical implication is then clearly
\be
{\rm Im}\,a_\rho(s)>0\,,\quad Y_\rho(s) \equiv 0\,.\label{eq: physical condition}
\ee
The first condition makes physical sense, as the spectral density has to be nonzero for nontrivial physics. We can also use the second condition to locate the physical bound state or resonance above the cut-off \cite{Caron-Huot:2021rmr,Albert:2022oes}. It is worth noting, however, that this is the most ideal situation. As previously mentioned, in practice, it is not possible to treat EFT bootstrap as an infinite dimensional SDP. Therefore, practically, we expect the duality gap is not zero but will converge to zero as we increase the dimension of the problem. In this situation, the weak duality can serve as a double check criteria, helping us diagnose any numerical mistakes. Practical implementation of EFT bootstrap also complicates the task of solely using $Y_\rho(s)\equiv 0$ to determine the physical points \cite{Albert:2022oes}. Nonetheless, we will demonstrate later that the first condition from \eqref{eq: physical condition} with sufficiently small $Y_\rho(s)$ can still be insightful for extracting physical information. 

Other scenarios of EFT bootstrap can be formulated similarly, we keep the discussions in appendix \ref{app: EFT}.

\section{Large-$N$ chiral perturbation theory}
\label{sec: review chPT}
Starting from this section and in all subsequent sections, we will apply the EFT bootstrap to large-$N$ $\chi$PT. We are adopting the set-up presented in \cite{Albert:2022oes}, where the dual algorithm was employed. 

Our innovations compared to \cite{Albert:2022oes,Fernandez:2022kzi} are twofold. For the dual algorithm, we use the crossing symmetric sum rules. As we previously indicated, these rules automatically incorporate all null constraints, making them more efficient and allowing us to easily explore higher dimensional operator; Additionally, we will establish the primal method and demonstrate the convergence between the primal and dual approaches.

\subsection{Chiral Lagrangian and low-energy amplitudes}

We consider the chiral limit of large-$N$ QCD (with SU$(N)$ gauge group)  \cite{t1993planar}, where the fermionic sector possesses ${\rm U}_L(N_f)\times {\rm U}_R(N_f)$ chiral symmetry. Usually, there is an axial anomaly which breaks the global symmetry by ${\rm U}_L(N_f)\times{\rm U}_R(N_f)\rightarrow {\rm SU}_L(N_f)\times{\rm SU}_R(N_f)\times {\rm U}_V(1)$. However, the axial anomaly is suppressed by the large-$N$ limit \cite{Witten:1979vv,Veneziano:1979ec,Kaiser:2000gs}.  At low-energy, we then have the spontaneous symmetry breaking pattern
\be
{\rm U}_L(N_f)\times {\rm U}_R(N_f) \rightarrow {\rm U}_V(N_f)\,.
\ee
This results in $N_f^2-1$ pseudo-Goldstone bosons (which are massless in the chiral limit) in the adjoint representation of SU$(N_f)$. There is also a singlet meson that can mix with the gluon. However, due to large-$N$ understanding of the OZI rule, this mixing is suppressed by  $1/N$, making it the trivial U$(1)$ part of U$(N_f)$. For $N_f=2$, the adjoint representation contains pions $\pi$; for $N_f=3$, the adjoint representation contains pions $\pi$, kaons $K$ and the eta $\eta$; while the singlet is referred to as the eta prime $\eta^\prime$. Nevertheless, we will show momentarily (see also \cite{Albert:2022oes}) that the unitarity constraints are independent of $N_f$ at the strict large-$N$ limit. We therefore follow \cite{Albert:2022oes} to refer to the Goldstone bosons to as the large-$N$ pion.

We can formulate the large-N pion physics using the coset construction of
\be
{\rm U}_L(N_f)\times{\rm U}_R(N_f)/{\rm U}_V(N_f)\,,
\ee
and then construct a non-linear sigma model by parameterizing the symmetry breaking in terms of low-energy field
\be
U=\exp\big[2i\fft{\Pi^a T^a}{f_\pi}\big]\,,
\ee
where $f_\pi$ is the pion decay constant that scales as $\sqrt{N}$ in the large-$N$ limit. Here  $\Pi^a$ denotes the large-$N$ pion. For example, for $N_f=3$ we have
\be
\Pi=\fft{1}{\sqrt{2}}\left(
\begin{array}{ccc}
 \frac{\eta }{\sqrt{6}}+\frac{1}{\sqrt{2}} & \pi ^+ & K^+ \\
 \pi ^- & \frac{\eta }{\sqrt{6}}-\frac{1}{\sqrt{2}} & K^0 \\
 K^- & K^0 & -\sqrt{\frac{2}{3}} \eta  \\
\end{array}
\right)+ \fft{1}{\sqrt{6}}\eta^\prime \mathbb{I}\,,
\ee
where $\mathbb{I}$ is a $3\times 3$ identity matrix. Then the chiral Lagrangian describing the EFT can be constructed \cite{Gasser:1983yg}. Up to $p^4$ at large-$N$ limit, it is generally given by \cite{gasser1985chiral} 
\be
\mathcal{L}_{\chi {\rm PT}}&=-\fft{f_\pi^2}{4}{\rm Tr}\big[\partial_\mu U^{\dagger} \partial^\mu U\big]+l_1 {\rm Tr}\big[(\partial_\mu U^{\dagger} \partial^\mu U )^2\big]+l_2 {\rm Tr}\big[\partial_\mu U^{\dagger} \partial_\nu U\partial^\mu U^{\dagger} \partial^\nu U\big] + \cdots\,.\label{eq: chiral Lag}
\ee
It is worth noting that we drop all sub-leading terms that are not single-trace because a flavor trace comes from a quark loop and thus also acquires a color trace \cite{gasser1985chiral,peris1995large}. This leaves us only two independent Wilson coefficients. Interestingly, for finite $N$ but $N_f=2$, there are also just two independent Wilson coefficients up to $p^4$; while for finite $N$ but $N_f=3$, there are three independent Wilson coefficients at this order. Since we only consider $2$-to-$2$ pion scattering, we then simply drop all background gauge fields (see \cite{Albert:2023jtd} for dual bootstrap including background gauge fields). At higher orders, it becomes quite challenging to enumerate a complete set of higher-dimensional operators without redundancy, especially when identities exist that allow trading one operator for another in  $N_f=2,3$. However, using global symmetry and Bose symmetry, one can easily write down tree-level amplitudes to any order in $p$.

It turns out to be useful to parametrize the amplitudes using the generators of U$(N_f)$ \cite{chivukula1993analyticity}
\be
\mathcal{M}_{ab}\,^{cd}=4\Big({\rm Tr}\big(T_aT_b T^c T^d\big)+{\rm Tr}\big(T_bT_a T^d T^c\big)\Big)\mathcal{M}(s,t) + \delta_{ab}\delta^{cd}\hat{\mathcal{M}}(s,t) + {\rm perm}\,,\label{eq: generator basis}
\ee
where it is obvious $\mathcal{M}(s,t)=\mathcal{M}(t,s),\hat{\mathcal{M}}(s,t)=\hat{\mathcal{M}}(t,s)$, making the Bose symmetry manifest. At large-$N$ limit, $\hat{\mathcal{M}}(s,t)$ is trivially zero because it can only be contributed by non-planar diagrams and is therefore suppressed (see, e.g., \cite{chivukula1993analyticity} for an explicit one-loop result). Using the permutation symmetry, one can easily construct tree-level amplitudes at low-energy up to any orders in $p$
\be
\mathcal{M}_{\rm low}(s,t)=\sum_{m=1}^{\infty}\sum_{n=1}^{[\fft{m}{2}]}g_{mn} (s^{m-n}t^n+s^n t^{m-n})\,,
\ee
where we have removed $p^0$ term as it is forbidden by the Adler's zero \cite{adler1965consistency}. The low-lying identification with the Lagrangian is \cite{Albert:2022oes}
\be
g_{10}=\fft{1}{2f_\pi^2}\,,\quad g_{20}=\fft{2(l_1+2l_2)}{f_\pi^4}\,,\quad g_{21}=\fft{4l_2}{f_\pi^4}\,.
\ee

\subsection{Partial waves and unitarity}

We now turn our attention to reviewing the partial waves of large-$N$ pion scattering. This can be understood by considering the amplitude $\mathcal{M}_{ab},^{cd}$ as a sum over the U$(N_f)$ irreducible representations that label the $\Pi^a\Pi^b\rightarrow X$ three-point vertices. Here, $X$ represents the intermediate states in the $\Pi^a\Pi^b\rightarrow \Pi^c\Pi^d$ scattering. Following the notation in \cite{BandaGuzman:2020wrz}, the representation theory behind this physical picture is given by\footnote{We factorize U$(N_f)$ as ${\rm U}(1)\times {\rm SU}(N_f)$, and therefore denote ${\rm adj}= (0,{\rm adj})$. We then follow the flavour structure analysis in \cite{BandaGuzman:2020wrz} by treating the $0$ component trivially.}
\be
{\rm adj} \otimes {\rm adj} = 0 \oplus {\rm adj}_{\rm S} \oplus {\rm adj}_{\rm A} \oplus \bar{a}s \oplus \bar{s}a \oplus \bar{s}s\oplus \bar{a}a\,.\label{eq: reps}
\ee
The multiplicity of $2$ for the adjoint representation arises because the vertices can be either symmetric or anti-symmetric in two legs. The resulting adjoint representation labels meson states consisting of bilinear quarks (i.e., $\bar{q}q$ states) in the intermediate channel, while the other representations label different exotic resonances. In addition to these global symmetries, $\Pi^a$ behaves like a scalar, making the partial waves trivially correspond to the Legendre polynomials, as in scalar scattering. Therefore, one has the following $s$-channel partial wave decomposition \cite{Albert:2022oes}
\be
\mathcal{M}_{ab}\,^{cd}(s|t,u)= \sum_{\mathcal{R}} \big(\mathbb{P}^{\mathcal{R}}\big)_{ab}\,^{cd}\, \mathcal{M}^{\mathcal{R}}(s|t,u)\,,\label{eq: irrep basis}
\ee
where
\be
\mathcal{M}^{\mathcal{R}}(s|t,u)= 16\pi \sum_J (2J+1)\, a_J^{\mathcal{R}}(s) P_J\big(1+\fft{2t}{s}\big)\,.
\ee
Here, $\mathcal{R}$ denotes the irreducible representations in \eqref{eq: reps}, and $\mathbb{P}^{\mathcal{R}}$ is the projector associated with $\mathcal{R}$. The crucial difference from the pure scalar case is that the Bose symmetry also permutes the projector $\mathbb{P}^{\mathcal{R}}$. The unitarity condition is then also straightforward
\be
\Big|1+i a_J^{\mathcal{R}}(s)\Big|^2\leq 1\,.\label{eq: unitarity pion}
\ee

The projector associated with the adjoint representation can be easily constructed because there are only two simple $\Pi^a\Pi^bX^c$ vertices: either the structure constant $f^{abc}$ or $d^{abc}:=2{\rm Tr}\big[T^a {T^b,T^c}\big]$. The two associated projectors are therefore proportional to $f^{abe}f^{ecd}$ and $d^{abe}d^{ecd}$. Other projectors are more intricate but can be constructed using the Casimir operators and the relevant eigenvalues \cite{BandaGuzman:2020wrz}. We document all projectors in appendix \ref{app: projector}. From the permutation symmetry of the projectors, we can easily write down the spin selection rules for \eqref{eq: unitarity pion} 
\be
0, {\rm adj}_{\rm S}, \bar{s}s, \bar{a}a: {\rm even}\, J\,,\quad {\rm adj}_{\rm A},  \bar{a}s\oplus\bar{s}a: {\rm odd}\, J\,.
\ee

To perform the EFT bootstrap, it is beneficial to explicitly know the relations between the generator basis \eqref{eq: generator basis} and the basis from irreducible representations \eqref{eq: irrep basis}. This can be readily achieved if we are aware of all the projectors, and we have
\be
& \mathcal{M}^{0}(s|t,u)=-\fft{2}{N_f}\mathcal{M}(t,u)+\fft{2(N_f^2-1)}{N_f}\big(\mathcal{M}(s,t)+\mathcal{M}(s,u)\big)+\hat{\mathcal{M}}(t,u)+\fft{N_f^2}{2}\big(\hat{\mathcal{M}}(s,t)+\hat{\mathcal{M}}(s,u)\big)\,,\nn\\
& \mathcal{M}^{{\rm adj}_{\rm S}}(s|t,u)=-\fft{4}{N_f}\mathcal{M}(t,u)+\fft{(N_f^2-4)}{N_f}\big(\mathcal{M}(s,t)+\mathcal{M}(s,u)\big)+\hat{\mathcal{M}}(t,u)+\fft{1}{2}\big(\hat{\mathcal{M}}(s,t)+\hat{\mathcal{M}}(s,u)\big)\,,\nn\\
&  \mathcal{M}^{{\rm adj}_{\rm A}}(s|t,u)=N_f\big(\mathcal{M}(s,t)-\mathcal{M}(s,u)\big)+\fft{1}{2}\big(\hat{\mathcal{M}}(s,t)-\hat{\mathcal{M}}(s,u)\big)\,,\nn\\
&  \mathcal{M}^{\bar{s}s}(s|t,u)=2\mathcal{M}(t,u)+\hat{\mathcal{M}}(t,u)+\fft{1}{2}\big(\hat{\mathcal{M}}(s,u)+\hat{\mathcal{M}}(s,t)\big)\,,\nn\\
&  \mathcal{M}^{\bar{a}a}(s|t,u)=-2\mathcal{M}(t,u)+\hat{\mathcal{M}}(t,u)+\fft{1}{2}\big(\hat{\mathcal{M}}(s,u)+\hat{\mathcal{M}}(s,t)\big)\,,\nn\\
& \mathcal{M}^{\bar{a}s\oplus\bar{s}a}(s|t,u)=\fft{1}{2}\big(\hat{\mathcal{M}}(s,t)-\hat{\mathcal{M}}(s,u)\big)\,.
\ee
After we take $\hat{\mathcal{M}}=0$ due to the large-$N$ limit, we can reproduce the relations outlined in \cite{Albert:2022oes}. Using these relations, we can easily translate the unitarity condition \eqref{eq: unitarity pion} into constraints on partial wave coefficients of $\mathcal{M}(s,t), \mathcal{M}(s,u)$ and $\mathcal{M}(t,u)$ 
\be
& \mathcal{M}(s,t)=16\pi \sum_J (2J+1) a_J^{st}(s) P_J\big(1+\fft{2t}{s}\big)\,,\quad \mathcal{M}(s,u)=16\pi \sum_J (2J+1) a_J^{su}(s) P_J\big(1+\fft{2t}{s}\big)\,,\nn\\
& \mathcal{M}(t,u)=16\pi \sum_{{\rm even}\,J} (2J+1) a_J^{tu}(s) P_J\big(1+\fft{2t}{s}\big)\,,
\ee
where we have $c_J^{su}=(-1)^J c_J^{st}$. This aids in constructing both the dual and primal problems. In the large-$N$ limit, it suffices to use the positivity bootstrap (this will be justified in the next subsection). Additionally, all exotic mesons are suppressed, and therefore ${\rm Im}\, c^{\bar{s}s}\equiv 0$. We then have \cite{Albert:2022oes}
\be
& {\rm Im}\, a_J^{st}(s) = {\rm Im}\, a_J^{su}(s)=\fft{N_f}{4(N_f^2-1)} {\rm Im}\, a^{0}(s) \geq 0\,,\quad a^{{\rm adj}_S}(s)= \fft{N_f^2-4}{2(N_f^2-1)} a^{0}(s)\,,\quad \text{for even}\, J\,,\nn\\
& {\rm Im}\, a_J^{st}(s) = -{\rm Im}\, a_J^{su}(s)=\fft{1}{2N_f} {\rm Im}\, a^{{\rm adj}_{\rm A}}(s) \geq 0\,,\quad \text{for odd}\, J\,,\nn\\
& {\rm Im}\,a^{tu}(s)\equiv 0\,.\label{eq: positivity condition}
\ee
It is obvious that, in the large-$N$ limit, the bootstrap constraints would be independent of the number of flavours. However, other bootstrap scenarios depend on $N_f$.

\subsection{Dual problem set-up}

Let's set up the dual problem for large-$N$ pion scattering. The most crucial component is the crossing symmetric dispersive sum rules. As we previously described, the construction of these sum rules is universal. The theory-dependent inputs involve constructing the crossing symmetric amplitudes and making assumptions about their Regge behaviour.

In QCD, the Regge intercept is at $J_0\simeq 0.52$ for both $\mathcal{M}(s,t)$ and $\mathcal{M}(s,u)$ \cite{Pelaez:2003ky,Pelaez:2004vs}, therefore $k_0=1$. We follow \cite{Albert:2022oes} to assume that the assumption $k_0=1$ remains valid after taking the large-$N$ limit. This improved Regge growth is usually assumed in QCD-like theories and SMEFT to constrain the low dimensional operators \cite{Low:2009di,Falkowski:2012vh,Bellazzini:2014waa}\footnote{We are grateful to Brando Bellazzini for pointing out the relevant references that we previously missed.}.

It turns out that one can construct three independent crossing symmetric amplitudes \cite{Zahed:2021fkp}
\be
& \mathcal{M}^{(1)}=\mathcal{M}(s,t)+\mathcal{M}(t,u)+\mathcal{M}(s,u)\,,\nn\\
& \mathcal{M}^{(2)} = \fft{\mathcal{M}(s,t)-\mathcal{M}(s,u)}{t-u} + \text{cyc perm}\,,\nn\\
& \mathcal{M}^{(3)} = \Big(\fft{\mathcal{M}(s,t)-\mathcal{M}(s,u)}{t-u} - \fft{\mathcal{M}(s,t)-\mathcal{M}(t,u)}{s-u}\Big)\fft{1}{s-t} + \text{cyc perm}\,,
\ee
To construct well-defined crossing symmetric sum rules from these amplitudes, we need to determine their Regge behaviors using the Regge boundedness of the building blocks $\mathcal{M}(s,t)$. We find
\be
k_0^{(1)}=1\,,\quad k_0^{(2)}=k_0^{(3)}=-1\,.
\ee
Thus, in terms of these symmetric amplitudes, the sum rules for $\mathcal{M}^{(2,3)}$ are super-convergence sum rules. The complete set of sum rules is therefore
\be
 B_k^{(1)}=\oint_{z=1,\xi,\xi^2}\fft{dz}{4\pi i}\mathcal{K}_{k+1}(z)\mathcal{M}^{(1)}(z,p^2)\equiv 0\,,\quad B_k^{(2,3)}=\oint_{z=1,\xi,\xi^2}\fft{dz}{4\pi i}\mathcal{K}_{k-1}(z)\mathcal{M}^{(2,3)}(z,p^2)\equiv 0\,,
\ee
where $k=1,3,5\cdots$, denoting the Regge spin of the sum rules with respect to $\mathcal{M}(s,t)$. The low-lying low-energy contributions from these sum rules are
\be
& -B_1\Big|_{\rm low}=\Big\{4g_{20}-2g_{21}+3p^2(2g_{30}-g_{31}),3g_{10},-3(g_{20}-2g_{21})\Big\}\,,\nn\\
&
-B_3\Big|_{\rm low}=\Big\{\left(6 g_{60}-3 \left(g_{61}+g_{62}-2 g_{63}\right)\right) p^4+\left(10 g_{50}-5 g_{51}+g_{52}\right) p^2+2 \left(2 g_{40}-g_{41}+g_{42}\right),\nn\\
& 3 \left(g_{40}+g_{41}-2 g_{42}\right) p^2+3 g_{30},-3 \left(g_{50}-2 g_{51}+g_{52}\right) p^2-3 \left(g_{40}-2 g_{41}+2 g_{42}\right)\Big\}\,.
\ee
As we noted, using these sum rules, we don't need to construct the null constraints as in \cite{Albert:2022oes,Fernandez:2022kzi,Albert:2023jtd}. At high energy, we have
\be
& B_k^{(1)}\Big|_{\rm high}=\Big\langle \left((-1)^J+1\right) m^{-3 k-5} \left(2 m^2+3 p^2\right) \left(m^2+p^2\right)^{\frac{k-1}{2}} P_J(x)\Big\rangle\,,\nn\\
& B_k^{(2)}\Big|_{\rm high}=\Big\langle \ft{3 m^{5-3 k} \left(\left((-1)^J+1\right) \sqrt{m^2+p^2} \sqrt{m^2-3 p^2}-\left(\left((-1)^J-1\right) m^2\right)-3 \left((-1)^J-1\right) p^2\right) \left(m^2+p^2\right)^{\frac{k-4}{2}}}{2 \sqrt{m^2-3 p^2}} P_J(x)\Big\rangle\,,\nn\\
& B_k^{(3)}\Big|_{\rm high}=\Big\langle \ft{3 m^{3-3 k} \left(-\left((-1)^J+1\right) \sqrt{m^2+p^2} \sqrt{m^2-3 p^2}+3 \left(1-(-1)^J\right) m^2+3 \left(1-(-1)^J\right) p^2\right) \left(m^2+p^2\right)^{\frac{k-4}{2}}}{2 \sqrt{m^2-3 p^2}}P_J(x) \Big\rangle\,,
\ee
where $x=\big((m^2-3p^2)/(m^2+p^2)\big)^{1/2}$. The average is defined by
\be
\Big\langle\cdots \Big\rangle = 8 \sum_J(2J+1) \int_{M^2}^{\infty} dm^2 \,{\rm Im}\,a^{st}(m^2) \big(\cdots\big)\,.
\ee

We can now decide what bootstrap scenario that we should use. Let's simply look at $B_1^{(2)}$, at leading order in $p^2\rightarrow 0$, we have
\be
1\gg g_{10}=\Big\langle \fft{1}{m^4}\Big\rangle> 0\,.\label{eq: g10>0}
\ee
This does not only prove the positivity of $g_{10}$ \cite{Albert:2022oes}, but it also satisfies the condition of using only positivity, because $g_{10}\sim 1/f_\pi^2\sim 1/N \ll 1$. Therefore, we will focus on the positivity bootstrap, with the exception of subsection \ref{subsec: upper}. In subsection \ref{subsec: upper}, we will employ the linear unitarity bootstrap to verify that the large-$N$ expansion is meaningful at the EFT level. We use SDPB \cite{Simmons-Duffin:2015qma,Landry:2019qug} to implement the algorithm.

\subsection{Primal problem set-up}

Let's now focus on the setup of the primal problem. The fundamental building block of the primal problem is the S-matrix ansatz, which approximates the S-matrix \cite{Paulos:2017fhb}. To ensure that we are indeed constructing a primal problem that is dual to the previously described dual problem, this ansatz must satisfy the assumptions of analyticity and Regge boundedness. Consequently, it has to validate all sum rules.
 
To ensure analyticity, we follow the approach in \cite{Paulos:2017fhb} to define a function using Mandelstam variables
\be
\rho_s=\fft{M-\sqrt{M^2-s}}{M+\sqrt{M^2-s}}\,.
\ee
 This function obviously has branch cut starting at $s=M^2$. Then the ansatz can be built by polynomials in $(\rho_s, \rho_t, \rho_u)$ under the restrictions of Bose symmetry and momentum conservation $s+t+u=0$.
 
Let's now use $\rho_{s,t,u}$ to construct the ansatz for $\mathcal{M}(s,t)$, which is symmetric in $s$ and $t$. It is crucial not to overcount or miss any terms in the ansatz. Since the primary problem is to determine the optimal coefficients of the ansatz terms, any redundancy or omission could either produce unfaithful bounds or simply disrupt the numerical calculations. We start with listing the generators of our ansatz that are symmetric in $(s,t)$
\be
(\rho_s \rho_t)^a\,,\quad (\rho_s+\rho_t)^b\,.
\ee
We do not need to consider $\rho_u^a$ because the large-$N$ limit suppresses the $u$-channel cut of $\mathcal{M}(s,t)$, as previously reviewed. Typically, the relation $s+t+u \equiv 0$ constrains the number of independent polynomials at each order, starting from order $5$, which need to be subtracted \cite{Paulos:2017fhb}. In our case, since there is no $\rho_u$ available in $\mathcal{M}(s,t)$ in the large-$N$ limit, the polynomials of the form $(s+t+u) \times (\cdots)$ do not exist. Hence, polynomials constructed using the above generators are independent. We can easily write down the ansatz\footnote{It is worthing noting that this ansatz actually has maximal analyticity, which is stronger than the analytic assumptions of the dispersive sum rules. Nevertheless, as we show below, the dual and primal bounds converge, seemingly suggesting that the SDP set-up of the positivity EFT bootstrap does not use the maximal analyticity. We are grateful to Miguel Correia for the discussions on this point.}
\be
\mathcal{M}(s,t)=\mathcal{R}(s,t)\sum_{a+b>0}^{2a+b=N_{\rm max}} \alpha_{ab}\, (\rho_s\rho_t)^a (\rho_s+\rho_t)^b\,,\label{eq: ansatz large-N}
\ee
where the lowest order is $1$ rather than $0$ due to the Adler's zero when expanding in $s,t\ll M^2$. The overall function $\mathcal{R}(s,t,u)$ is for controlling the Regge behaviour of the amplitudes. Because $k_0=1$, for simplicity, we choose
\be
\mathcal{R}(s,t)=1\,.
\ee
We can modify $\mathcal{R}(s,t)$ to exhibit a more refined Regge behaviour by specifying $J_0$. However, since the only ingredient needed for constructing the solutions of the positivity EFT bootstrap comes from the sum rules, and these sum rules are sensitive to $k_0$ rather than $J_0$, it's natural to speculate that modifying $\mathcal{R}(s,t)$ won't alter the resulting bounds as long as it grows at a rate below $s^{k_0=1}$ at high energy. We will verify this point in subsection \ref{subsec: Regge}.

By expanding the ansatz \eqref{eq: ansatz large-N} in the low-energy limit where $s,t,u\ll M^2$, we can derive the low-energy tree-level amplitudes and establish a dictionary that translates Wilson coefficients to $\alpha_{ab}$. For example, the dictionary for low-lying coefficients is
\be
g_{10}=\fft{1}{4}\alpha_{10}\,,\quad g_{20}=\fft{1}{16}(2\alpha_{01}+\alpha_{02})\,,\quad g_{30}=\fft{1}{64}(5\alpha_{01}+4\alpha_{02}+\alpha_{03})\,,\quad g_{21}=\fft{1}{32}(2\alpha_{02}+\alpha_{10})\,.\label{eq: dic low-lying}
\ee
We can then easily verify that every term in \eqref{eq: ansatz large-N} satisfies the crossing symmetric sum rules.

The primal problem involves imposing the positivity condition \eqref{eq: positivity condition} on the ansatz \eqref{eq: ansatz large-N} and solving for the coefficients $\alpha_{ab}$ by optimizing the targeted Wilson coefficients using the dictionary like \eqref{eq: dic low-lying}. The last technical question is how we read off the partial wave coefficients from our ansatz \eqref{eq: ansatz large-N}? We use the standard inversion formula\footnote{This formula may explain why the primal bootstrap typically does not employ maximal analyticity, thus converging to the dual one with weaker analyticity. For $s \geq M^2$, this formula solely relies on the physical regime $-M^2 \leq t < 0$, making the resulting data sensitive only to the analyticity for $t < 0$. We, therefore, propose examining the subtlety of 'maximal analyticity vs. partial analyticity' using the gravitational EFT. In this context, unitarity must be demanded beyond integer spin, e.g., for $J \sim b \sqrt{s}$ at high energy \cite{Caron-Huot:2021rmr,Caron-Huot:2022ugt}.}
\be
a_J(s)=\frac{2^{3-2 d} \pi ^{1-\frac{d}{2}}}{\Gamma \left(\frac{d}{2}-1\right)}s^{\fft{d-4}{2}} \int _{-1}^1dx (1-x)^{\fft{d-4}{2}} \mathcal{M}(s,t)P_J(x)\,,\quad x= 1+\fft{2t}{s}\,.
\ee
It is crucial to note that, although we only consider the imaginary part in the positivity SDP Lagrangian, we can still solve the full S-matrix from the optimal primal solutions. While it may seem that the primal method always provides more information than the dual, this perception is mistaken. On the dual side, one can also rely on the extreme functional to employ the analytic ``rule-in'' method, which enables the construction of a relevant UV theory \cite{Caron-Huot:2020cmc,Chiang:2021ziz,Albert:2022oes}. We use SDPB \cite{Simmons-Duffin:2015qma,Landry:2019qug} to implement the algorithm.

\section{Dual bounds meet primal dounds}
\label{sec: bounds}
\subsection{Simple linear bounds}

\subsubsection{``Trivial" positivity bounds}

Let us start with positivity bounds of $g_{10}, g_{20}$ and $g_{21}$. The Lagrangian, as explicitly written down, are
\be
& \mathcal{L}_{1} = g_{10}- \int_{M^2}^{\infty}ds \sum_J {\rm Im}\,a_J(s) Y^1_J(s)\,,\quad \mathcal{L}_{2} = g_{20}+\lambda_{2}(g_{10}-1)- \int_{M^2}^{\infty}ds \sum_J {\rm Im}\,a_J(s) Y^{2}_J(s)\,,\nn\\
& \mathcal{L}_{3} = g_{21}+\lambda_{3}(g_{10}-1)- \int_{M^2}^{\infty}ds \sum_J {\rm Im}\,a_J(s) Y^3_J(s)\,.\label{eq: simple Lag}
\ee
Using the crossing symmetric sum rules, it is easy to obtain the dual bounds, we have \eqref{eq: g10>0} as well as
\be
&\fft{1}{3}\big(B_1^{(1)}+\fft{1}{3}B_1^{(3)}\big)\Big|_{p=0} \rightarrow g_{20}=\Big\langle \fft{1}{m^6}\Big\rangle> 0\,,\nn\\
&\fft{1}{6}\big(B_1^{(1)}+\fft{4}{3}B_1^{(3)}\big)\Big|_{p=0} \rightarrow g_{21}=\Big\langle \fft{1-(-1)^J}{m^6}\Big\rangle \geq 0\,.
\ee

The primal bounds for $g_{10}, g_{20}>0$ are also trivial to obtain, where the solutions are all $\alpha_{ab}\equiv 0$, since $g_{10}=0$ or $g_{20}=0$ would a trivial free theory. This is consistent with the Slater 's condition and the complementary condition previously reviewed: the dual functional $Y_J(s)$ is strictly positive, therefore the strong duality gives ${\rm Im}\,a_J(s)\equiv 0$. 

\begin{figure}[h]
\centering
\includegraphics[width=.6\linewidth]{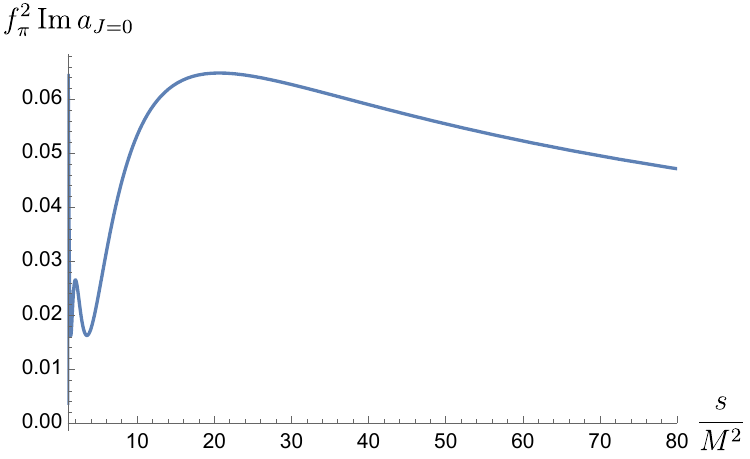} 
\caption{The spectral density at $J=0$ that saturates $\tilde{g}_{21}=0$, where the S-matrix is approximated by $N_{\rm max}=5$ polynomial.}\label{fig: g21low}
\end{figure}

The first nontrivial example is $g_{21}$, since its dual functional $Y_J(s)$ can be zero for even spins, suggesting that nontrivial UV amplitudes with only even spin particles exist. It turns out that $g_{21}\geq 0$ converges trivially for a low $N_{\rm max}=5$ and a low $J_{\rm max}$, which we choose to be $J_{\rm max}=60$. A nontrivial S-matrix profile with $J=0$ that saturates $g_{21}=0$ can be illustrated, as shown in Fig \ref{fig: g21low}. We observed that the spectral density for all higher spins is zero. This preliminary study thus confirms the statement from \cite{Albert:2022oes} that the UV theory at $\tilde{g}_{21}=0$ is a scalar theory. However, we observe from Fig \ref{fig: g21low} that the UV scalar spectral density doesn't show an extreme peak at certain points; instead, it presents a continuum. This suggests that the UV theory is not a single scalar but a scalar theory with all possible mass values where $m\geq M$.

\subsubsection{Upper bounds on $g_{20}/g_{10}$ and $g_{21}/g_{10}$}

To bootstrap the upper bounds, we simply flip the overall sign of $g_{20}, g_{21}$ in the Lagrangians \eqref{eq: simple Lag}. The upper bound of $g_{20}$ in the unit of $g_{10}$ is also trivialized by the dual method 
\be
\fft{1}{3}\big(-B_1^{(1)}+B_1^{(2)}+\fft{1}{3}B_1^{(3)}\big)\Big|_{p=0}\rightarrow g_{10}-g_{20}=\Big\langle \fft{m^2-M^2}{m^6}\Big\rangle \geq 0 \rightarrow \tilde{g}_{2}=g_{20}/g_{10}M^2\leq 1\,.\label{eq: upbound g20}
\ee
The upper bound of $\tilde{g}_{2}^\prime=2g_{21}/g_{10}M^2$, although it is not straightforward, it can still be easily solved from the dual algorithm using SDPB \cite{Simmons-Duffin:2015qma}. Using the crossing symmetric sum rules, we reproduced the result of \cite{Albert:2022oes}
\be
\tilde{g}_{2}^\prime\leq 3.25889135\,.
\ee

\begin{figure}[h]
\begin{subfigure}{.4\textwidth}
\centering
\includegraphics[height=0.23\textheight]{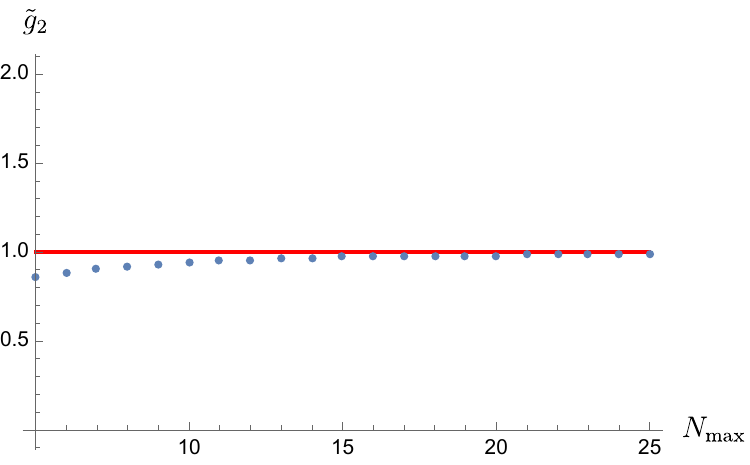} 
\caption{}
\label{subfig: pp2}
\end{subfigure}
\hfill
\begin{subfigure}{.4\textwidth}
\centering
\includegraphics[height=0.23\textheight]{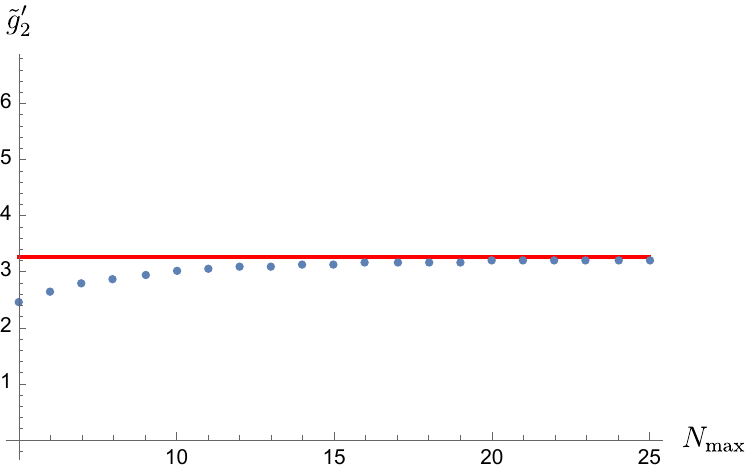}  
\caption{}
\label{subfig: pm2}
\end{subfigure}
\caption{We determined the upper primal bounds of $\tilde{g}_2$ and $\tilde{g}_2^\prime$ by varying $N_{\rm max}$. These bounds converge quickly to the dual bounds, which are represented by red lines, approaching from below as guaranteed by the strong duality of SDP.
}
\label{fig: simple upper bounds}
\end{figure}

These two bounds are nontrivial from the primal side, since low spin sampling and low $N_{\rm max}$ would give us trash, which does not extrapolate well to an infinite dimensional SDP. For a more involved S-matrix bootstrap, which involves either linear unitarity or even complete unitarity, the strategy is to fix $N_{\rm max}$ and then increase $J_{\rm max}$ so that one can extrapolate the bounds to be valid for all $J$; subsequently, one should vary $N_{\rm max}$ and extrapolate the bounds to $N_{\rm max}=\infty$ \cite{Guerrieri:2021ivu,Guerrieri:2022sod}. However, for the positivity primal bootstrap, we find that we can simply fix $J_{\rm max}$ to a large value without doing the extrapolation. We choose $J_{\rm max}=60$, and we can see the nice convergence of bounds by varying $N_{\rm max}$ from $5$ to $25$, as shown in Fig \ref{fig: simple upper bounds}. When $N_{\rm max}$ takes a small value, the approximation of the positivity EFT SDP is not good. However, we still expect the weak duality to be valid. This is precisely why we see that the primal upper bounds are always smaller than the dual upper bounds. Ultimately, we find that $N_{\rm max}=25$ is enough to conclude the strong duality, as the relative error of primal bounds from the dual bounds is roughly $\sim 1\%$.

Now we can combine the dual and primal functionals to analyze the physical spectrum that saturates bounds. For primal side, we simply use the solutions from $N_{\rm max}=25$. The crucial point to understand is that since we are still far from the actual infinite-dimensional SDP, we cannot rely exclusively on either the dual functional or the primal solution to extract physical information. The strategy is as follows: initially, examine the dual functional. If the dual functional is precisely zero at a particular point, then we should trust the spectral density from the primal solution at that point, irrespective of its magnitude size. Conversely, if the dual functional is strictly positive and large, we would expect the corresponding primal ``spectrum'' to be small. Ideally, this primal spectrum should be vanishingly small. If it's not, it should be small enough to be interpreted as a numerical artifact, and we should simply discard it. The most subtle situation arises when the dual functional is strictly positive and small enough to be approximated as zero. In this case, we should estimate the gap $Y_J(s){\rm Im}\,a_J(s)$ at that point. If the gap is sufficiently small, we can trust the primal spectrum; otherwise, we discard the data. However, this is also difficult to implement. It is worth noting that the dual functional is usually a polynomial of $M^2/s$, and is small for sufficiently high $s$ numerically. It is challenging to numerically detect that a small number is a zero or it is simply suppressed by $1/s$. This suggests that the EFT bootstrap does not have sharp implications in the deep UV, but a vague picture of the physics there can still be captured by the primal solutions: for large $s$, we believe that it is reasonable to trust the primal spectrum density, because we can always treat $Y_\rho(s)$ there as zero with small errors.

\begin{itemize}
 \item $\tilde{g}_2=1$
 \end{itemize}

\begin{figure}[h]
\centering
\includegraphics[width=.6\linewidth]{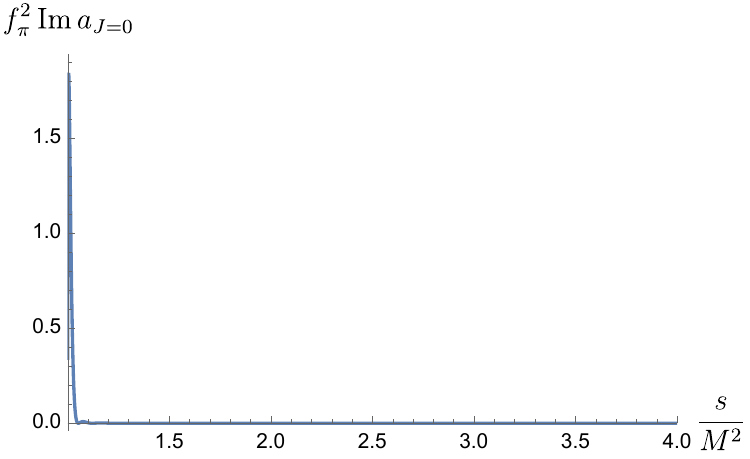} 
\caption{The spectral density at $J=0$ that saturates $\tilde{g}_{2}=0$. To generate this plot, we used a polynomial order in S-matrix of $N_{\rm max}=25$.}\label{fig: g20upspec}
\end{figure}

\begin{figure}[h]
\begin{subfigure}{.4\textwidth}
\centering
\includegraphics[height=0.18\textheight]{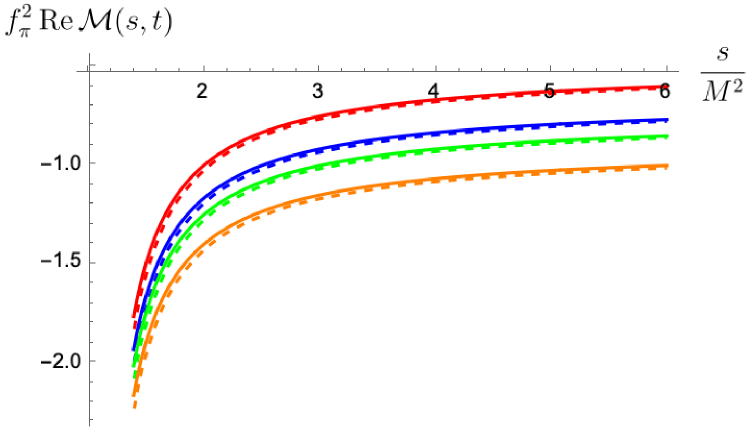} 
\caption{}
\label{subfig: amp1}
\end{subfigure}
\hfill
\begin{subfigure}{.5\textwidth}
\centering
\includegraphics[height=0.18\textheight]{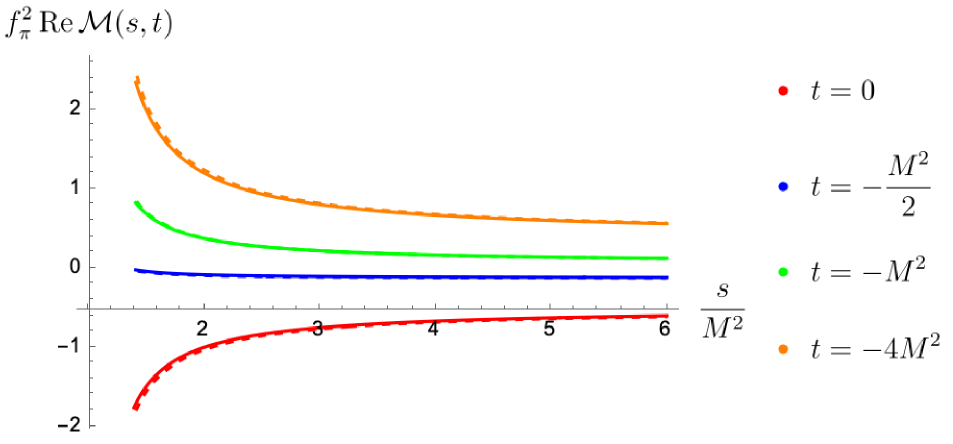}  
\caption{}
\label{subfig: amp2}
\end{subfigure}
\caption{(a) The real part of the amplitudes $\mathcal{M}(s,t)$ with $\tilde{g}_2=1$ for several values of $t$. (b) The real part of the amplitudes $\mathcal{M}(s,t)$ with $\tilde{g}_2^\prime\simeq 3.26$ for several values of $t$. The solid lines are drawn from the analytic rule-in amplitudes \eqref{eq: analytic-rule-in g20} and \eqref{eq: analytic-rule-in g21} respectively, while the dashed lines are drawn from the primal solutions.
}
\label{fig: g20 and g21 upamp}
\end{figure}

From the dual functional \eqref{eq: upbound g20}, we see that $Y_J(s)$ can be zero only when $s=M^2$ and it is strictly positive for $s>M^2$, which is robust against adding more functionals. We indeed observe from the primal solution that there is a single peak around $s=M^2$ for $J=0$, see Fig \ref{fig: g20upspec}; while ${\rm Im}\,a_{J}$ for $J\geq 1$ is vanishing. Besides, we also checked that the other Wilson coefficients at this point are $\tilde{g}_{2}^\prime=0$ and $\tilde{g}_3=g_{30}/g_{10}M^4\simeq 0.976\sim 1$. This analysis confirms that the UV theory with $\tilde{g}_2=1$ corresponds to a single scalar theory with mass $m=M$, as first pointed out by \cite{Albert:2022oes}. The relevant scalar mode with $s^0$ Regge behaviour is
\be
\mathcal{M}_{\rm scalar}(s,t)= \fft{M^2}{2f_\pi^2}\big(\fft{s}{M^2-s}+\fft{t}{M^2-t}\big)\,.\label{eq: analytic-rule-in g20}
\ee
A comparison of this scalar amplitudes with our numerical solution is illustrated in Fig \ref{subfig: amp1}. Nevertheless, it is important to note that the plot in Fig. \ref{subfig: amp1} is drawn for the global region, which significantly suppresses the differences between the analytic and numerical amplitudes. The largest difference between amplitudes obtained by the two methods occurs at $|s|\rightarrow\infty$ and $t\rightarrow0$, and is approximately $0.018$. 

\begin{itemize}
\item $\tilde{g}_2^\prime\simeq 3.25889135$
  \end{itemize}

\begin{figure}
    \centering
    \begin{subfigure}{0.45\textwidth}
        \includegraphics[width=\linewidth]{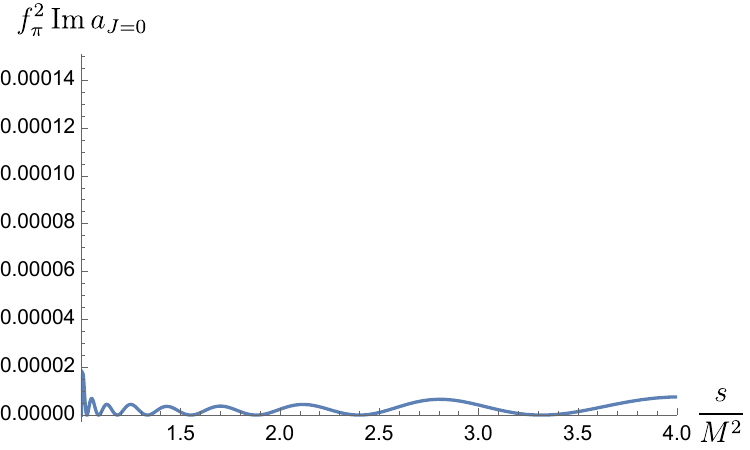}
        \caption{}
        \label{fig:sub1}
    \end{subfigure}
    \hfill
    \begin{subfigure}{0.45\textwidth}
        \includegraphics[width=\linewidth]{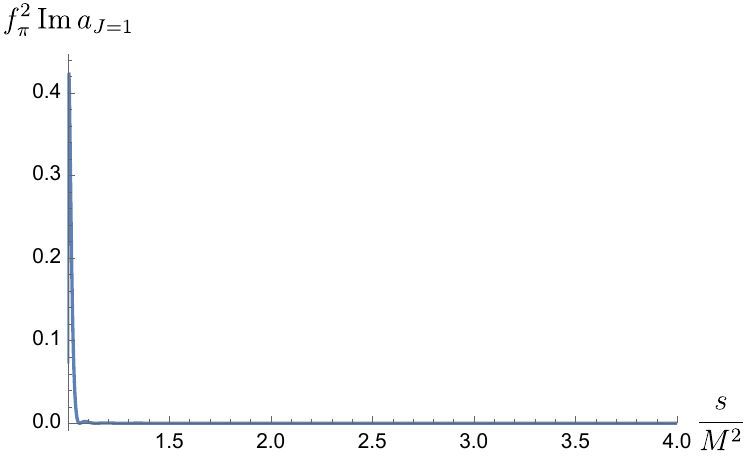}
        \caption{}
        \label{fig:sub2}
    \end{subfigure}
    \vspace{1em} 
    \begin{subfigure}{0.45\textwidth}
        \includegraphics[width=\linewidth]{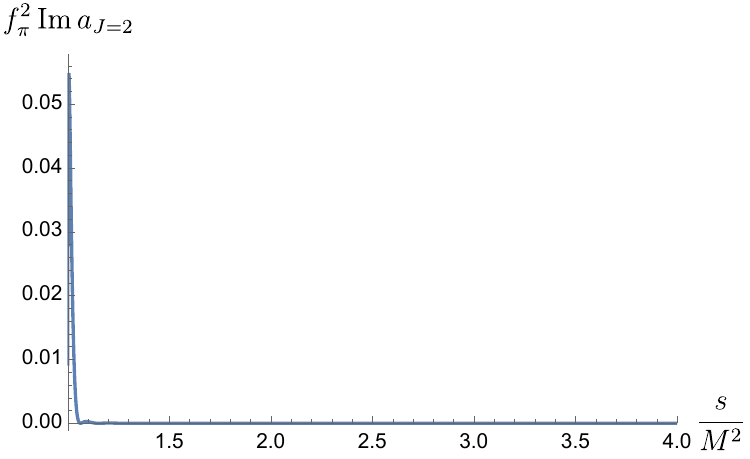}
        \caption{}
        \label{fig:sub3}
    \end{subfigure}
    \hfill
    \begin{subfigure}{0.45\textwidth}
        \includegraphics[width=\linewidth]{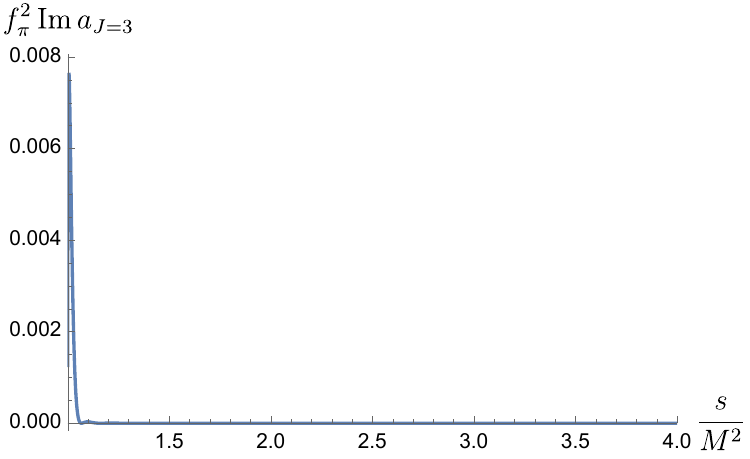}
        \caption{}
        \label{fig:sub4}
    \end{subfigure}
    \caption{The spectral density from $J=0$ to $J=3$ is solved using $N_{\rm max}=25$ at $\tilde{g}_2^\prime\simeq 3.26$. The $J=0$ spectral density \label{fig:sub1} is nonphysical, as it is of the order $10^{-5}$, which is parametrically small compared to others. For higher $J$ values, we have smaller spectral densities, serving as a reminder of the low-spin dominance.}
    \label{fig: g21upspectrum}
\end{figure}

The associated dual functional is complicated, but we can nevertheless easily observe that it is strictly positive for $J=0$ but it is zero for $J>0, s=M^2$. From the primal side, we indeed observe that ${\rm Im}\,a_{J=0}$, although not exactly zero, is parametrically small as of order $10^{-5}$; in addition, $J>0$ spectral density exhibits a sharp pump around $s=M^2$, which is, however, getting smaller and smaller for larger $J$. See Fig \ref{fig: g21upspectrum} for an illustration with $J=0,1,2,3$. 
In addition, we find that this point gives $\tilde{g}_2\simeq 0.99\sim 1$ and $\tilde{g}_3\simeq 0.97\sim 1$. This analysis confirms the statement of \cite{Albert:2022oes} that the UV theory with $\tilde{g}_2^\prime\simeq 3.26$ is a theory with $J\geq 1$ and $m=M$. However, it is important to note that our ``numerical theory'' is radically different from the $su$-model which also saturates $\tilde{g}_2^\prime\sim 3.26$ \cite{Albert:2022oes}, because the $su$-model has Regge behaviour $s^{-1}$ while our ansatz grows like $s^0$. We can modify the $su$-model to describe a spin-$1$ theory with Regge behaviour $s^0$
\be
\mathcal{M}_{su-{\rm mod}}(s,t)=\fft{M^2}{2(1-\log 2)f_\pi^2}\Big[\fft{s\, t}{(M^2-s)(M^2-t)}+(1-\log 2)\Big(\fft{s}{M^2-s}+\fft{t}{M^2-t}\Big)\Big]\,.\label{eq: analytic-rule-in g21}
\ee
We can then compare our numerical rule-in with this analytic rule-in in Fig. \ref{subfig: amp2}, where the maximal difference is around $0.017$ 

\subsubsection{The Skyrme bound and a mysterious Regge trajectory}

There is an interesting linear bound, giving rise to precisely the Skyrme model \cite{makhankov2012skyrme,lenz1997lectures}, as first noticed in \cite{Albert:2022oes}. This bound involves $g_{21}$ and $g_{20}$, and we can formulate it as
\be
\mathcal{L}=-g_{21}+\lambda(g_{20}-1) - \int_{M^2}^\infty ds \sum_J {\rm Im}\,a_J(s) Y_J(s)\,.
\ee
The bound is again trivial on the dual side, we have
\be
B_1^{(1)}(p=0)\rightarrow 4g_{20}-2g_{21}=\Big\langle \fft{2(1+(-1)^J)}{s^3}\Big\rangle \geq 0 \rightarrow \tilde{g}_2^\prime \leq 4 \tilde{g}_2\,.
\ee
According to the previous experience, $N_{\rm max}=25$ is good enough for us to perform a nice primal algorithm. We then find that the primal bound is
\be
\tilde{g}_2^\prime \lesssim  3.939\, \tilde{g}_2\,,\quad \text{primal bound with $N_{\rm max}=25$}\,.
\ee
The error from the dual rigorous bound is around $1.54\%$.

\begin{figure}[h]
    \centering
    \begin{subfigure}{0.45\textwidth}
        \includegraphics[width=\linewidth]{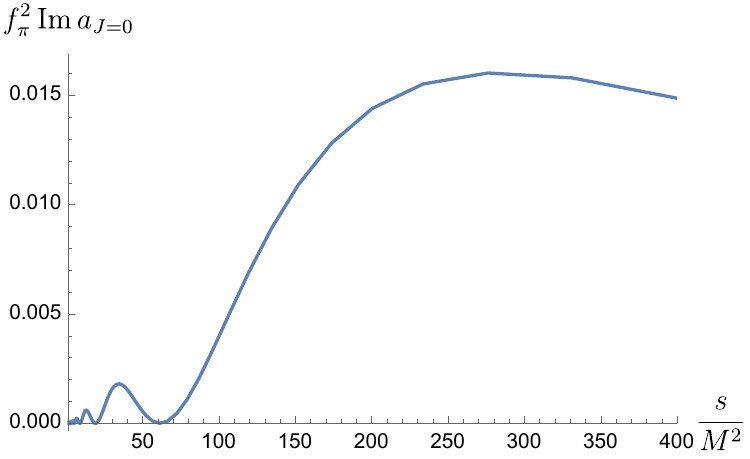}
        \caption{}
        \label{fig:skyrmesub1}
    \end{subfigure}
    \hfill
    \begin{subfigure}{0.45\textwidth}
        \includegraphics[width=\linewidth]{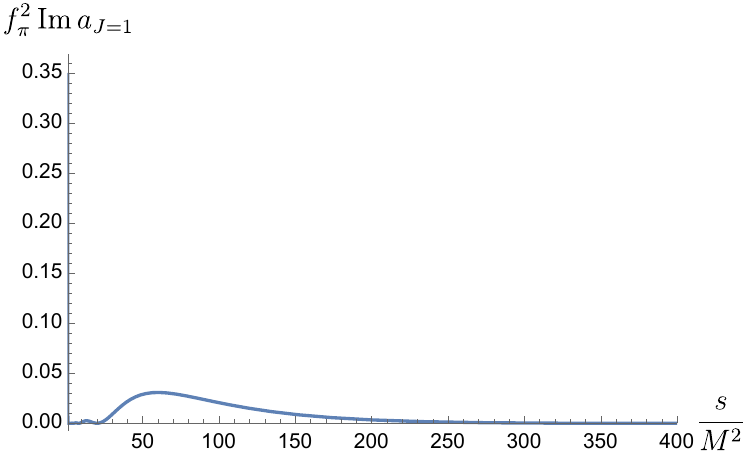}
        \caption{}
        \label{fig:skyrmesub2}
    \end{subfigure}
    \vspace{1em} 
    \begin{subfigure}{0.45\textwidth}
        \includegraphics[width=\linewidth]{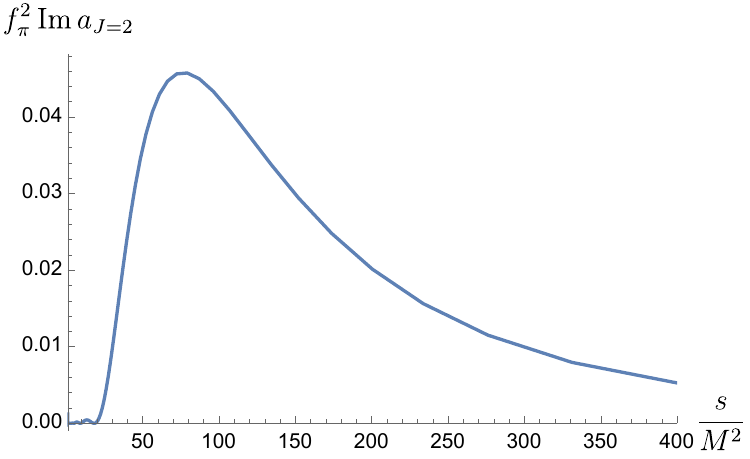}
        \caption{}
        \label{fig:skyrmesub3}
    \end{subfigure}
    \hfill
    \begin{subfigure}{0.45\textwidth}
        \includegraphics[width=\linewidth]{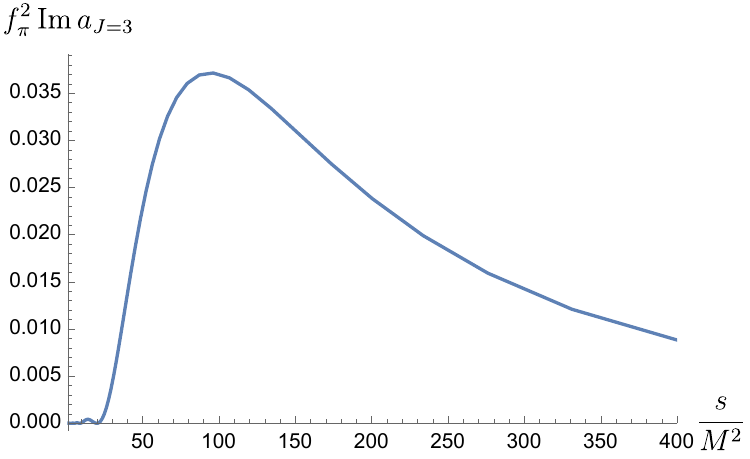}
        \caption{}
        \label{fig:skyrmesub3}
    \end{subfigure}
    \caption{The spectral density for $J=0,1,2,3$ is solved using $N_{\rm max}=25$ on the Skryme line $\tilde{g}_2^\prime=4\tilde{g}_2$. The vector meson dominance is exhibited. For $J=1$, there is a vector $\rho$ meson with mass around $M$ and also other resonance around $7.4M$ that might be a numerical flaw; for other spins, there are dominated resonances at energy $s>50M^2$.}
    \label{fig: skyrmespectrum}
\end{figure}

Let's now turn to analyze the physical spectrum of the Skyrme model. We observe that the simple dual functional $Y_J(s)$ can be zero only for odd $J$, which seems to suggest that we should discard all data with even spin in the primal solution. However, this conclusion is not robust against expanding the space of the dual functionals. When using $126$ functionals, we can find that the dual functional is small at $J=0,s=M^2$ (around $10^{-3}$), while it is strictly positive and not small close to $s=M^2$ for other $J$. With this in mind, we can examine the primal solution, and we find that the contributions from $J\neq 1$ are relatively smaller than those from $J=1$. This behaviour can be described as the vector meson dominance \cite{sakurai1960theory}. Specifically, for $J=1$ we identify a sharp peak around $m=M$, which can be interpreted as a vector $\rho$ meson. There is also significant physics at higher energies: a continuum with a resonant bump around $m \approx 7.4 M$ and a width of roughly $12.6M$, which might be a numerical artifact when compared to the $m=M$ peak. For other spins, we should only trust the behaviour at sufficiently high $s=m^2$, and we indeed observe dominate resonances at energies $m > 7 M$ with a relatively wide width. It is important to note that these resonances are stable upon increasing $N_{\rm max}$. Interestingly, if we consider $M$ as the mass of the $\rho$ meson, approximately $770{\rm MeV}$, then the mass of all those heavy resonances exceeds $4000{\rm MeV}$ and thus would contain, e.g., a bottom quark. See Fig \ref{fig: skyrmespectrum} for an explicit illustration for $J=0,1,2,3$. 

\begin{figure}[h]
\centering
\includegraphics[width=.6\linewidth]{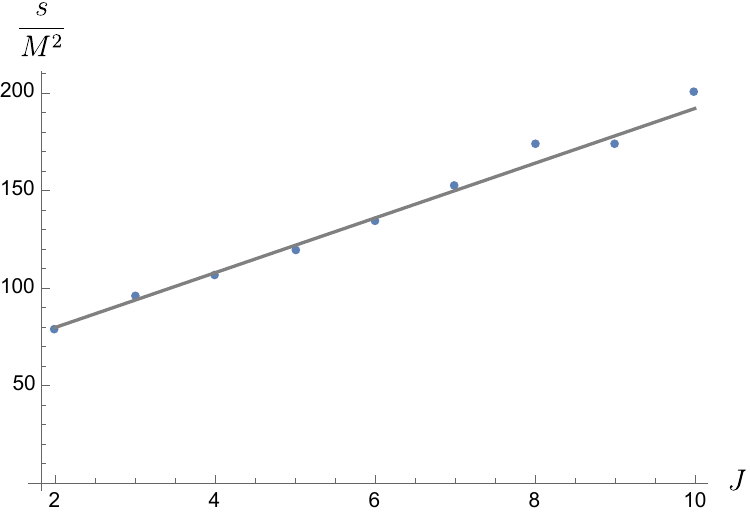} 
\caption{The Regge trajectory is read off from the primal solutions. This trajectory can be fitted as a linear line with minor errors.}
\label{fig: Regge}
\end{figure}

We can extend our primal analysis up to $J=10$ and find subsequent resonances. More surprisingly, these resonances can be organized as an approximately linear Regge trajectory\footnote{We are grateful to Gabriel Cuomo and Victor Rodriguez for suggesting this interesting exercise.}, as shown in Fig \ref{fig: Regge}. For higher $J$, we observe a significant deviation from the fitted Regge trajectory. This deviation is likely due to poor numerical shooting. Unfortunately, we have no clear explanation for this trajectory. It is possible that we should not take it from such high energy behaviour of primal solutions, and as inferred by \cite{Albert:2022oes}, those resonances are actually pushed to infinity in the large-$N$ limit. Refining the numerics to better understand this Regge trajectory would be interesting in the future.

\begin{figure}[h]
\centering
\includegraphics[width=.65\linewidth]{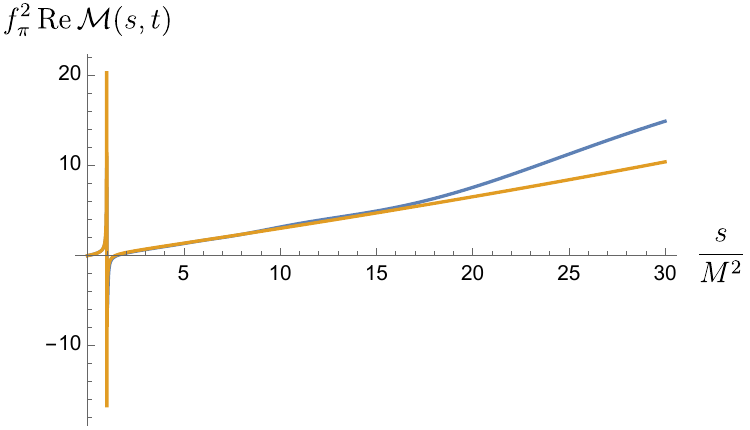} 
\caption{The comparison between the numerical amplitude from the primal solution with $N_{\rm max}=25$  (blue) and the UV model \eqref{eq: UV model} with $m_{\infty}\simeq 14.7M$ (yellow).}
\label{fig: Skyrmeamp}
\end{figure}

Now we can compare the numerical amplitude to the model proposed in \cite{Albert:2022oes}
\be
\mathcal{M}_{{\rm spin}-1}^{(\rm UV)} = \fft{M^2}{2f_\pi^2} \fft{m_{\infty}^2}{M^2+3m_{\infty}^2}\Big(\fft{M^2+2t}{M^2-s} \fft{m_{\infty}^2}{m_{\infty}^2-t}+\fft{M^2+2s}{M^2-t} \fft{m_{\infty}^2}{m_{\infty}^2-t} \Big)\,,\label{eq: UV model}
\ee
where it reduces to a single $\rho$ meson model in the limit $m_\infty\rightarrow\infty$. Our strategy of comparison is to first solve $m_{\infty}$ for requiring the Regge limit of this analytic model equals the numerical amplitude at fixed $t$ and then compare the amplitudes with other energy. We take $t=-1/10$ and find that $m_{\infty}\big|_{t=-1/10}\simeq 14.7 M$, the comparison is displayed in Fig \ref{fig: Skyrmeamp}. We observe that although the extremely high energy limit is required to be the same, two amplitudes become clearly distinguishable at $s\sim 15 M^2$. The difference at large $s$ but below $|s|\rightarrow\infty$ is anticipated, because \eqref{eq: UV model} is just a toy model with a single resonance $m_{\infty}$ to adjust, while the primal solutions Fig \ref{fig: skyrmespectrum} contain more than one resonance, organized as a Regge trajectory Fig \ref{fig: Regge}. We can also evaluate the Wilson coefficients that saturate the Skyrme bound from our primal solution, $\tilde{g}_2\simeq 0.364, \tilde{g}_2^\prime\simeq 1.434, \tilde{g}_3\simeq 0.349$, which are close to what \eqref{eq: UV model} predicts $(1/3,4/3,1/3)$.

\subsection{Exclusion plots}

\subsubsection{$\mathcal{O}(p^4)$}

So far, we have only dealt with simple linear bounds. The power of the EFT bootstrap lies in its ability to search for allowed spaces that involve multiple Wilson coefficients. These represent nonlinear bounds, implying that the boundary is not a linear function. Let's focus on the space spanned by $\tilde{g}_{2}^\prime$ and $\tilde{g}_2$ and reproduce the exclusion plot made using dual methods in \cite{Albert:2022oes}.

Our strategy is to search in different directions in the $\tilde{g}_{2}^\prime-\tilde{g}_2$ plane. The corresponding Lagrangian is
\be
\mathcal{L}_{p^4}= \big(\cos (2\pi c) g_{20} + \sin (2\pi c) g_{21}\big)+\lambda (g_{10}-1) - \int_{M^2}^\infty ds \sum_J {\rm Im}\,a_J(s) Y_J(s)\,,
\ee
where $c\in [0,1)$. On the dual side, this corresponds to fixing the objective and varying the normalization condition; while on the primal side, this precisely means fixing the normalization to $g_{10}=1$ and bounding different linear combinations of $g_{20}$ and $g_{21}$. There is another approach, different from these angle-searching methods. In this approach, one can fix a particular value of one parameter, for example, $g_{20}/g_{10}$, and search for the upper and lower bounds of another parameter $g_{21}/g_{10}$ \cite{Caron-Huot:2020cmc}. While this method is typically as efficient as the previous one on the dual side, it complicates the search on the primal side. This is because it fixes two parameters, $g_{20}$ and $g_{10}$, making the positive matrices degenerate. Consequently, an additional transformation is needed to generate a valid positivity input. Nevertheless, near the corner, it is aways better to adopt the ``fixing-parameter'' method rather than the ``angle-searching''. We leave the details of ``fixing-parameter'' methods in appendix \ref{app: fixing-parameter}.

\begin{figure}[h]
\centering
\includegraphics[width=.7\linewidth]{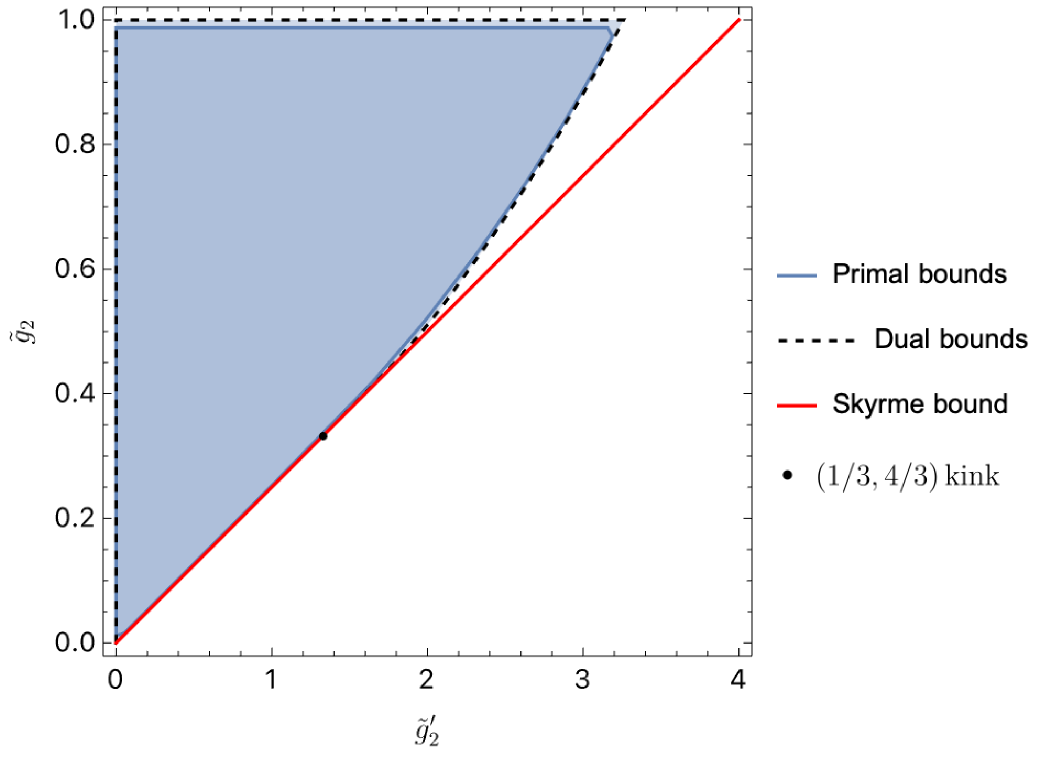} 
\caption{The exclusion plot involves $\tilde{g}_2$ and $\tilde{g}_2^\prime$, and compares the primal bounds to the dual bounds. The red line represents the linear Skyrme bound and the black dot is the position of $(1/3,4/3)$ kink.}
\label{fig: g20g21plot}
\end{figure}

By sampling a sufficient number of $c$ values (roughly $100$ points) and few ``fixing-parameter'' points near the corner, we can make a sufficiently nice exclusion plot using both dual and primal methods (where we choose $N_{\rm max}=25$). The plot is displayed in Fig \ref{fig: g20g21plot}, where the dashed black boundary is drawn using the dual method, coinciding with the findings in \cite{Albert:2022oes}; on the other hand, the solid boundary is derived from the primal method. We find that the primal bounds efficiently converge to the dual bounds. To generate the dual bounds, we use $126$ functionals that are constructed from the crossing symmetric sum rules $B_1, B_3, B_5$\footnote{In the completion stage of this paper, \cite{McPeak:2023wmq} appeared and showed that assuming ${\rm Im}\,a^{tu}=0$ in complex scalar yields the same bound as the large-$N$ pion bounds.}.

We observe from Fig \ref{fig: g20g21plot} that the Skyrme line represents only a small segment of the entire boundary. The point where the bounds begin to deviate from the Skyrme model is referred to as a ``kink'' in \cite{Albert:2022oes}. Nontrivial physics is anticipated at this kink \cite{El-Showk:2012cjh,Albert:2022oes}. Further studies suggest that the kink is located at the point $(\tilde{g}_2,\tilde{g}_2^\prime)=(1/3,4/3)$ \cite{Fernandez:2022kzi}, which is ruled-in by the analytic model \eqref{eq: UV model} with $m_{\infty}\rightarrow \infty$. We present several points on the boundary in Fig \ref{fig: g20g21plot}, retaining three digits after the decimal, as shown in Table \ref{tab: table points}. We note that around $\tilde{g}_{2}\simeq 0.26$, the value of $\tilde{g}_2^\prime$ is already slightly smaller than what the Skyrme bound predicts, which is, however, likely to be numerical error. Therefore, using our dual numerical results does not provide sensible way to clearly pinpoint the position of the kink. 

\begin{table}[h]
\centering
\begin{tabular}{|c|c|}
\hline
$\tilde{g}_2$ & $\tilde{g}_2^\prime$ \\ 
\hline
$0.001$ & $0.004$  \\
$0.005$ & $0.020$ \\
$0.020$ & $0.080$ \\
$0.120$ & $0.480$ \\
$0.250$ & $1.000$ \\
$0.260$ & $1.039$ \\
$0.333$ & $1.332$ \\
$0.390$ & $1.577$ \\
\hline
\end{tabular}
\caption{Few boundary points with small $\tilde{g}_{2}$, obtained using the dual methods.}
\label{tab: table points}
\end{table}

\subsubsection{$\mathcal{O}(p^6)$}

\begin{figure}[h]
\centering
\includegraphics[width=.7\linewidth]{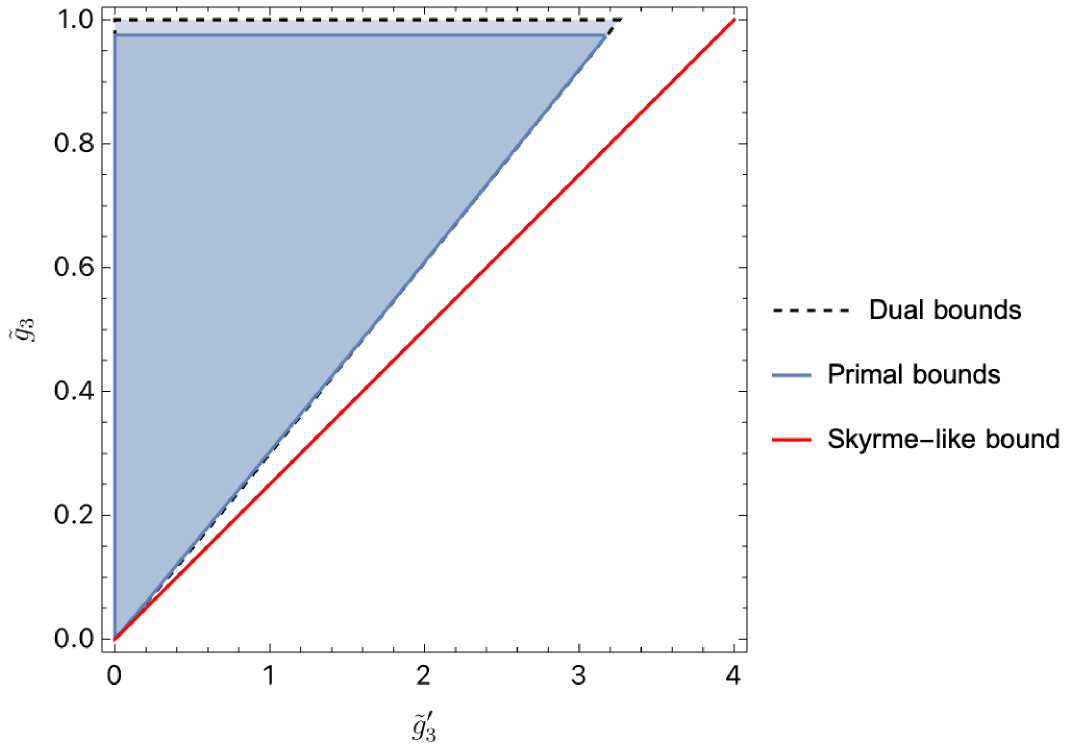} 
\caption{The exclusion plot involves $\tilde{g}_3$ and $\tilde{g}_3^\prime$, with comparison of the primal bounds to the dual bounds. The red line represents the linear Skyrme-like bound $\tilde{g}_3^\prime\leq 4 \tilde{g}_3$.}
\label{fig: g30g31plot}
\end{figure}

For $\mathcal{O}(p^6)$, we consider a similar Lagrangian but for $g_{30}, g_{31}$
\be
\mathcal{L}_{p^6}= \big(\cos (2\pi c) g_{30} + \sin (2\pi c) g_{31}\big)+\lambda (g_{10}-1) - \int_{M^2}^\infty ds \sum_J {\rm Im}\,a_J(s) Y_J(s)\,,
\ee
Following the same strategy as noted previously, we can create the exclusion plot for $\tilde{g}_3=g_{30}M^4/g_{10}$ and $\tilde{g}_3^\prime=g_{31}M^4/g_{10}$. There is a good convergence between the primal and dual methods, as seen in Fig \ref{fig: g30g31plot}.

Some comments are in order. Interestingly, we find that the linear upper bound of $\tilde{g}_3$ and $\tilde{g}_3^\prime$ is the same as in $\tilde{g}_2$ and $\tilde{g}_2^\prime$. This is because that the corner can still be analytically ruled in by \eqref{eq: analytic-rule-in g21}. and we also have the Skyrme-like linear bound $\tilde{g}_3^\prime\leq 4 \tilde{g}_3$. However, it is obvious from Fig \ref{fig: g30g31plot} that the whole allowed region is much smaller than the region enclosed by the linear bounds, even though there is no clear kink. In this case, the nonlinearity largely shrinks the allowed space of EFT. The same phenomenon was also observed in gravitational EFT \cite{Bern:2021ppb,Caron-Huot:2022ugt,Chiang:2022jep}.

\subsection{Is the positivity primal bootstrap sensitive to the Regge behaviour?}
\label{subsec: Regge}

In this subsection, we aim to address the question of whether the Regge behaviour in the primal ansatz affects the primal bounds. The answer should be ``No" for the positivity bootstrap, as long as the Regge behaviour is below $k_0=1$ ensuring that our dual set-up remains valid. This is confirmed by the duality between the dual and primal methods: the dual bounds are only sensitive to $k_0$, which provides all sum rules. Therefore, the primal bounds should not depend on the specifics of the Regge behaviour of the amplitudes, as long as their growth rate is below $k_0$. However, if the Regge behaviour of the primal ansatz reaches $k_0$ or exceeds it, it cannot measure and provide bounds on the Wilson coefficients with Regge spin $k_0$.

Our strategy is to choose $\mathcal{R}(s,t)$ in the ansatz \eqref{eq: simple Lag} as
\be
\mathcal{R}(s,t)=\big((1+\sqrt{1-s})(1+\sqrt{1-t})\big)^{2J^0}\,.
\ee
For $J^0=0$, we recover the case that we have been studying. In general, this factor modifies the dictionary that relates the ansatz parameters $\alpha_{ab}$ to the low-energy Wilson coefficients, but it preserves the analyticity and grows as $s^{J^0}$ in the Regge limit. We will study the primal upper bound of $\tilde{g}_2^\prime$, which can be measured by $B_1$ sum rules and is sufficiently nontrivial. We will vary $J^0$ by taking several values: $J^0=(0,0.5,0.75,1,1.1)$ and observe how the primal bounds change accordingly. This exploration is displayed in Fig \ref{fig: g21upmodRegge} below. In Fig \ref{fig: g21upmodRegge}, we present only the results for $J^0=(0,0.5,0.75)$, and it's clear that they all converge to the dual rigorous bound at sufficiently large $N_{\rm max}$. For $J^0=1, 1.1$, as expected, we do not obtain any bounds.

\begin{figure}[h]
\centering
\includegraphics[width=.7\linewidth]{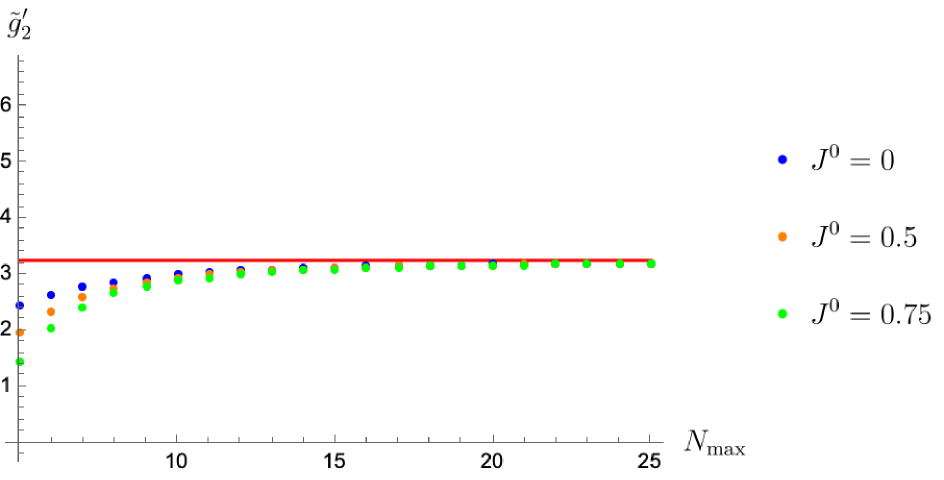} 
\caption{The primal upper bound of $\tilde{g}_{2}^\prime$ from the primal ansatz with different Regge behaviour $s^{J^0}$.}
\label{fig: g21upmodRegge}
\end{figure}

\subsection{An ad hoc: primally confirming the large-$N$ assumption}
\label{subsec: upper}

So far, we have been considering the constraints of the large-$N$ $\chi$PT, as suggested in \cite{Albert:2022oes}. As previously noted, we assume the large-$N$ limit from IR to UV; therefore, the positivity bootstrap is sufficient and strongly constraining. In this way, we can bound the Wilson coefficients in terms of $g_{10}=1/(2f_\pi^2)\sim 1/N$. Nevertheless, it is essential to question whether this assumption is justified from a low-energy point of view. In other words, do the large-$N$ bounds fall into the allowed regime of a more complete unitary region like linear unitarity?

This question may seem trivial, since there is no doubt that the spectral density scaling as $1/N$ falls into the linear unitarity ${\rm Im}\, a_J\sim 1/N \ll 2$. Indeed, this simple argument trivializes the linear bound: the large-$N$ limit yields bounds that scale in $g_{10}$, i.e., $g\leq \mathcal{O}(1) g_{10}/M^{{\rm dim}-2}$, which are significantly stronger than the bounds provided by linear unitarity, $g\leq \mathcal{O}(1)/M^{\rm dim}$, because $g_{10}\ll 1/M^2$. However, it is important to emphasize that this straightforward argument and power counting do not obviously work for nonlinear bounds. The dimensional analysis can only infer that the large-$N$ exclusion plot resides in the near-zero corner in the linear unitarity plot, however, its boundary may bend outside of the region of the linear unitarity exclusion plot. This concern arises due to results of \cite{Chiang:2022ltp,Chen:2022nym}, where it turns out that the boundary of the linear unitarity bounds is more curved and is sandwiched between the linear bounds.

\begin{figure}[h]
\centering
\includegraphics[width=.7\linewidth]{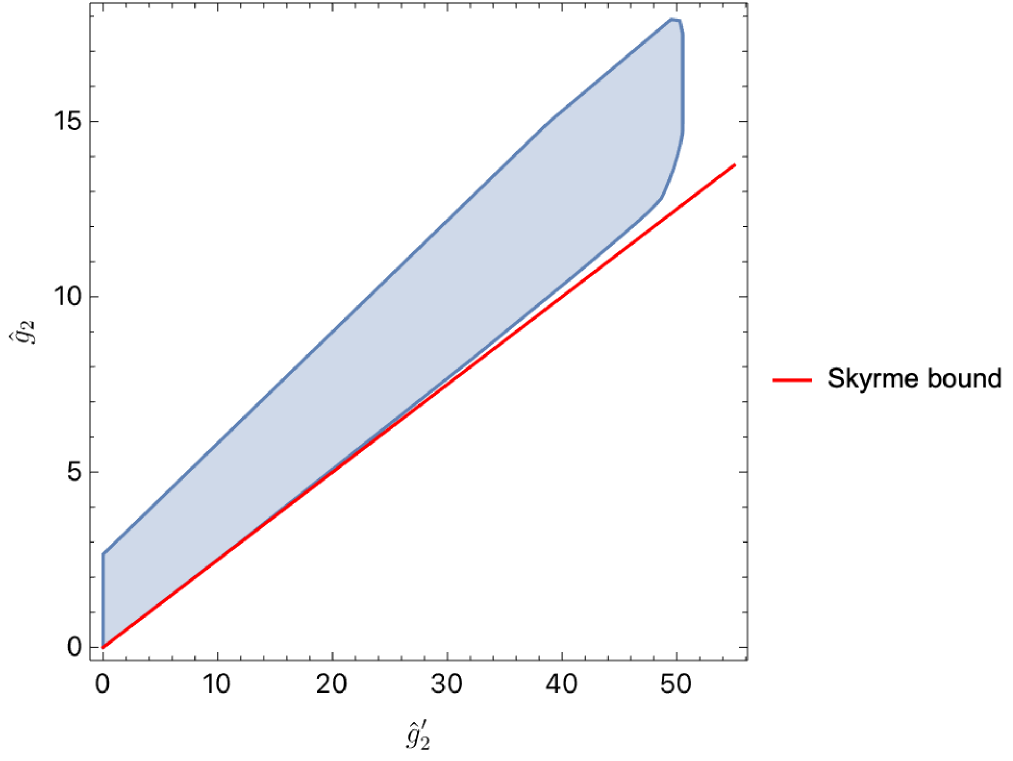} 
\caption{The exclusion plot for $\hat{g}_2$ and $\hat{g}_2^\prime$ from the primal linear unitarity bootstrap.The red line still represents the linear Skyrme bound $\hat{g}_2^\prime\leq 4 \hat{g}_2$.}
\label{fig: g20g21plotup}
\end{figure}

Our strategy to justify the large-$N$ bound involves using the linear unitarity bootstrap for pion amplitudes, whose structures remain constrained by the large-$N$ limit. We employ only the primal method. For the dual method that incorporates the upper bound of unitarity, see \cite{Caron-Huot:2020cmc,Chiang:2022ltp,Chiang:2022jep,Chen:2023bhu}. For a recent systematic numerical algorithm on the dual side, refer to \cite{Chen:2023bhu}\footnote{Interestingly, although the method in \cite{Chen:2023bhu} has a dual spirit, i.e., using the dispersion relation, the numerical algorithm seems to differ from SDP.}. For simplicity, we take $N_f=2$, then \eqref{eq: positivity condition} indicates that the unitarity constraints are
\be
0 \leq {\rm Im}\, a_{{\rm even}\,J}(s)\leq \fft{1}{3}\,,\quad 0 \leq {\rm Im}\, a_{{\rm odd}\,J}(s)\leq \fft{1}{2}\,.
\ee
Follow appendix \ref{app: EFT}, we then consider the following SDP Lagrangian
\be
\mathcal{L}_{\rm up} & =\big(\cos(2\pi c)g_{20}+\sin(2\pi c)g_{21}\big)  -\int_{M^2}^{\infty}ds\sum_{J} {\rm Im}\, a_J(s) Y_J(s)\nn\\
& - \int_{M^2}^{\infty}ds\sum_{{\rm even}\, J} \big(\fft{1}{2}-{\rm Im}\, a_J(s)\big) \tilde{Y}_J(s) -\int_{M^2}^{\infty}ds\sum_{{\rm odd}\, J} \big(\fft{1}{3}-{\rm Im}\, a_J(s)\big) \tilde{Y}_J(s)\,.
\ee
In this case, we physically ``normalize'' the free theory part $S=1+i T$. Although in this case, the numerical convergence is more slow, we find that $N_{\rm max}=25$ and $J_{\rm max}=60$ still suffice for our purpose. By searching different values of $c$, we found Fig \ref{fig: g20g21plotup}, where we define $\hat{g}_2^\prime =2 g_{21} M^4$ and $\hat{g}_2=g_{20} M^4$.

From Fig \ref{fig: g20g21plotup}, we can confirm that the linear bounds are indeed consistent with the large-$N$ bounds as long as $N\sim \mathcal{O}(10)$. Nevertheless, we do observe dangerous regions. The first dangerous region is around $\hat{g}_2 \sim 2.7$, where there is a sharp kink, and the boundary shrinks away from the vertical line. In contrast, the large-$N$ bounds in Fig \ref{fig: g20g21plot} do not exhibit this shrinking. To resolve this danger, we need to require
\be
f_\pi^2 \geq 0.19 M^2\,.\label{eq: cond safe}
\ee
The second danger is the Skyrme line. In this plot, the lower boundary behaves similarly to the large-$N$ plot: it first coincides with the Skyrme bound and then bends inward. Therefore, the position of the ``Skyrme kink'' in Fig \ref{fig: g20g21plotup} must be larger than the one in the large-$N$ case. Fortunately, the kink in Fig \ref{fig: g20g21plot} is around $\hat{g}_2^\prime \sim 11.6$, for which the condition \eqref{eq: cond safe} easily resolves the danger. Since \eqref{eq: cond safe} can be easily satisfied in the large-$N$ limit where $f_\pi^2\sim N\rightarrow\infty$, we thus confirm that assuming the large-$N$ limit at low energy is valid without paradox. Besides, we emphasize that this exercise also shows that the unitarity can be used to bound the decay constant in terms of the EFT scale $M$. Indeed, we can set  an EFT bootstrap to directly constrain the decay constant $f_\pi$ which deserves further exploration for finite $N$ $\chi$PT

\section{Constrain holographic QCD models}
\label{sec: holographic QCD}

Large-$N$ QCD enjoys holographic descriptions, which utilize the dual gravity theory to capture the salient properties of QCD, such as hadrons, in the strong coupling limit. One famous example of these models is known as the Witten-Sakai-Sugimoto model \cite{Witten:1998zw,Sakai:2004cn,Sakai:2005yt}, which can be constructed from string theory. Such models can also be built from low-energy EFTs of gauge fields with gravitational couplings \cite{Son:2003et,Kim:2009qs,Erlich:2005qh,DaRold:2005mxj,Hirn:2005nr,Panico:2007qd,Colangelo:2012ipa}, where the fits to the experimental data of hadrons, glueballs, and so on have been extensively studied (see e.g., \cite{Erdmenger:2007cm,Pahlavani:2014dma} for brief reviews). Typically, at low energy, the holographic models can also give rise to the $\chi$PT Lagrangian with Wilson coefficients mapping to parameters on the gravity side \cite{Sakai:2004cn,Sakai:2005yt,Kim:2009qs,Erlich:2005qh,DaRold:2005mxj,Hirn:2005nr,Panico:2007qd,Colangelo:2012ipa}. Therefore, we expect that the bounds on large-$N$ $\chi$PT can be translated into constraints on those holographic QCD models, carving out the allowed space of EFTs for gauge theories that can be consistently UV completed with gravity.

However, to our knowledge, all holographic QCD models so far only include ${\rm Tr}\, F^2$ term when deriving the $\chi$PT, giving rise to the Skryme model at order $\mathcal{O}(p^4)$. The essential reason is that higher derivative terms are relatively small, like suppressed by the string scale \cite{Sakai:2004cn,Sakai:2005yt}. Therefore, one can always adjust the fundamental parameters so that the Wilson coefficients live on the boundary of the exclusion plot \ref{fig: g20g21plot} below the kink. Nevertheless, although the higher derivative terms only give small corrections, these small corrections may still deform the Wilson coefficients outside of the allowed region in Fig \ref{fig: g20g21plot}. In this section, we will show that higher derivative terms ${\rm Tr}\, F^3$ and ${\rm Tr}\,F^4$ cause the low-energy theory deviate from the Skyrme model\footnote{Similarly, there exists a top-down $D_3-\bar{D}_5-\bar{D}_7$ construction of thermal QCD \cite{Mia:2009wj}, which, as including the $R^4$ term, was shown to give rise to the chiral Lagrangian beyond the Skryme model and compatible with the phenomenological data \cite{Yadav:2017bbe,Yadav:2020pmk}.}. Therefore, requiring the consistency with Fig \ref{fig: g20g21plot} puts constraints on the higher derivative couplings.

\subsection{Chiral Lagrangian from holographic QCD}

\subsubsection{Bulk theory and the power counting}

We now move to derive the chiral Lagrangian from $5D$ EFT of ${\rm SU}_{\rm L}(N_f)\times {\rm SU}_{\rm R}(N_f)$ gauge fields. We consider the following effective action with the background field $\bar{g}$

\be
S_{\rm eff}= \sum_{i={\rm L}, {\rm R}}\fft{1}{g_{\rm YM}^2}\int d^{5}x\sqrt{-\bar{g}}\,\varphi(x)\Big(-\fft{1}{2}{\rm Tr}\, F_i^2 + \fft{i g_H}{3}{\rm Tr}\, F_i^{(3)} + \fft{\alpha_1}{4} {\rm Tr}\, F_i^4+ \fft{\alpha_2}{4}{\rm Tr}\, F_i^{(4)}\Big)\,,\label{eq: Lag YM}
\ee
where
\be
& {\rm Tr}\,F^2={\rm Tr}\Big(F_{AB}F^{AB}\Big)\,,\quad {\rm Tr}\, F^{(3)}={\rm Tr}\Big(F_{A}\,^B [F_B\,^{C}, F_{C}\,^A]\Big)\,,\nn\\
&{\rm Tr}\,F^{4}={\rm Tr}\Big(F_{AB}F^{AB}F_{CD}F^{CD}\Big)\,,\quad  {\rm Tr}\,F^{(4)}={\rm Tr}\Big(F_{A}\,^B F_B\,^C F_C\,^D F_D\,^A\Big)\,.
\ee
All $A,B,\cdots$ refer to the five dimensional indices, and all indices so far are contracted by the background field $\bar{g}_{AB}$ and a background dilaton $\varphi$
\be
ds^2=\bar{g}_{AB}dX^A dX^B=b(z)^2 dz^2 +a(z)^2 \eta_{\mu\nu}x^\mu x^\nu\,,\quad z\in [z_{\rm UV},z_{\rm IR}]\,,\label{eq: background}
\ee
where we put an UV Randall-Sundrum (RS) bane \cite{Randall:1999ee,Randall:1999vf} at $z_{\rm UV}$ and an IR RS brane at $z_{\rm IR}$. For Anti de-Sitter (AdS) space, the UV brane is served as the boundary of AdS. Nevertheless, throughout this subsection, we consider $a(z), b(z)$ and $\varphi$ to be arbitrary. Their precise forms satisfy the equations of motion for both the gravity sector and the matter sectors (such as dilaton associated with $\varphi$ \cite{Sakai:2004cn,Sakai:2005yt,Karch:2006pv}). However, we treat the flavour gauge fields as probes \cite{Sakai:2004cn,Sakai:2005yt}. Appropriate boundary conditions are imposed on both the UV brane at $z_{\rm UV} \rightarrow 0$ and the IR brane $z_{\rm IR}$ to obtain the background solution \eqref{eq: background}. We ignore fluctuations from the graviton and other matters, as they are irrelevant for our purposes. 

How should we think about the EFT power counting of \eqref{eq: Lag YM} without a top-down picture like string theory? The bulk gauge fields are expected to correspond to conserved currents in QCD. Hence, we then have $J_L^{\mu a}$ and $J_R^{\mu a}$, which can be constructed by quark bilinear operators $J^{\mu a} = q \gamma^\mu T^a \bar{q}$; their conservation precisely reflects the global chiral symmetry. From large-$N$ counting, we know that we usually scale such quark bilinear operators by $1/\sqrt{N}$ to normalize the two-point function \cite{t1993planar}; this corresponds to scaling $A$ by $g_{\rm YM}$ to normalize the kinematic term, suggesting $g_{\rm YM}\sim 1/\sqrt{N}$ in \eqref{eq: Lag YM}. Moreover, the Wilson coefficients $g_H, \alpha_i$ should be suppressed by some EFT cut-off $\Lambda$, which could be the mass of higher spin particles or the string states (it's important to distinguish it from the $\chi$PT EFT cut-off $M$ for the moment). However, due to the presence of $g_{\rm YM}$, there are different possible consistent schemes of power counting for $g_H$ and $\alpha_i$, as long as their sizes don't grow beyond $\langle F\rangle^\#$. For instance, we can have $g_H\sim 1/\Lambda^{2}, \alpha_i \sim1/\Lambda^{4}$; or we can have $g_H\sim g_{\rm YM}^2/\Lambda, \alpha_i \sim g_{\rm YM}^2/\Lambda^{2}$, both of which are then suppressed by the large-$N$ limit. The idea is then to use the EFT constraints to decide the correct size of those Wilson coefficients. It's important to note that we actually ignore some double trace terms like $\big({\rm Tr}\,F^2\big)^2$, because ${\rm Tr}$ maps to the trace in $\chi$PT, and coefficients of double-trace operators in large-$N$ $\chi$PT are $1/N$ suppressed, for general $N_f$. Therefore, we conclude for $N_f>3$, without doing anything, that the Wilson coefficient of $1/g_{\rm YM}^2 \big({\rm Tr}\,F^2\big)^2$ must scale at least as $g_{\rm YM}^2/\Lambda^2$! Indeed, in the Witten-Sakai-Sugimoto model from the $D_4$-$D_8$-$\bar{D}_8$ intersection, the effective action is the dilaton-Born-Infeld (DBI) action \cite{Sakai:2004cn}, which still has only a single trace when generalized to the non-Abelian case  \cite{Hagiwara:1981my}.

\subsubsection{Routine to obtain the chiral Lagrangian}

We follow the strategy of \cite{Panico:2007qd} that uses the IR boundary condition to break the chiral symmetry
\be
A_{L\mu}(z_{\rm IR})-A_{R\mu}(z_{\rm IR})=0\,,\quad F_{L\mu z}(z_{\rm IR})+F_{R\mu z}(z_{\rm IR})=0\,.
\ee
This implies that the chiral symmetry breaks to its vector subgroup at IR. For convenience, we regroup the gauge group by its vector component and axial component
\be
\mathcal{V}_\mu = \fft{1}{2}(A_{L\mu} + A_{R\mu})\,,\quad \mathcal{A}_\mu= \fft{1}{2}(A_{L\mu} - A_{R\mu})\,.
\ee
It turns out that one can define a Wilson line stretch from the IR point into the bulk
\be
U=P\Big\{e^{i\int_{z_{\rm IR}}^{z} dz A_{Lz}}e^{-i\int_{z_{\rm IR}}^{z} dz A_{Rz}}\Big\}\,,
\ee
which has the property of the pion field under the gauge transformation $U\rightarrow g_R U g_L^{-1}$, and thus it corresponds to $U$ in $\chi$PT at the UV point. For simplicity, we choose the gauge $A_{R z}\equiv 0$, and use $U$ to gauge away $A_{L z}$ using the gauge fixing prescription of \cite{Sakai:2004cn,Hirn:2005nr,Panico:2007qd}. This procedure allows us to define the following boundary condition
\be
& {\rm IR}:\,\quad \partial_z \mathcal{V}_\mu(x,z_{\rm IR})=0\,,\quad \mathcal{A}_\mu(x,z_{\rm IR})=0\,,\nn\\
& {\rm UV}:\,,\quad \mathcal{V}_\mu(x,z_{\rm UV})=\mathcal{A}_\mu(x,z_{\rm UV})= \fft{i}{2} U \partial_\mu U^\dagger\,.
\ee

To obtain an effective action, one can solve the bulk gauge fields with respect to these boundary conditions, and then substitute the solutions back to have the on-shell action. For simplicity, in this paper, we only focus on those terms up to $\mathcal{O}(p^4)$, therefore we can simply solve the equation of motion at leading order \cite{Panico:2007qd}
\be
\mathcal{V}_\mu(x,z)=\mathcal{V}_\mu(x,z_{\rm IR})+\cdots\,,\quad \mathcal{A}_\mu(x,z)=f_A(z)\mathcal{A}_\mu(x,z_{\rm IR})+\cdots\,,
\ee
where $\cdots$ refer to those terms contributing to higher orders like $\mathcal{O}(p^6)$, and one has 
\be
f_A(z)=c_A\int_{z}^{z_{\rm IR}} d\xi\fft{b(\xi)}{a(\xi)^2 \varphi(\xi)}\,,\quad f_A(z_{\rm UV})=1\,.
\ee
It is then straightforward to obtain the chiral Lagrangian \eqref{eq: chiral Lag}, where the pion decay constant $f_\pi$ and other Wilson coefficients are given by
\be
& l_1=\fft{1}{16 g_{\rm YM}^2}\int_{z_{\rm UV}}^{z_{\rm IR}} dz\frac{\varphi (z) \left(-4 b(z)^2 g_H \left(f_A(z){}^2-1\right) f_A'(z){}^2-2 b(z)^4 \left(f_A(z){}^2-1\right){}^2+\left(2 \alpha _1+\alpha _2\right) f_A'(z){}^4\right)}{b(z)^3}\,,\nn\\
& l_2=\fft{1}{8 g_{\rm YM}^2}\int_{z_{\rm UV}}^{z_{\rm IR}} dz\frac{\varphi (z) \left(f_A(z){}^2-1\right) \left(b(z)^2 \left(f_A(z){}^2-1\right)+2 g_H f_A'(z){}^2\right)}{b(z)}\,,\nn\\
& f_\pi^2=\fft{2}{g_{\rm YM}^2}\int_{z_{\rm UV}}^{z_{\rm IR}} dz\, \fft{a(z) \varphi(z)f_A'(z)^2}{b(z)}\,.\label{eq: dictionary}
\ee

\subsection{Bounds for different models}

\subsubsection{Witten-Sakai-Sugimoto model is healthy}

The Witten-Sakai-Sugimoto model, constructed from the $D_4$-$D_8$-$\bar{D}_8$ brane configuration in type IIA string theory, is considered a top-down model and is expected to be robust. Therefore, before imposing constraints on more general bottom-up models like \eqref{eq: Lag YM}, we aim to verify the health of the Witten-Sakai-Sugimoto model from a low-energy perspective.

The essential idea is to start with the $D_8$ brane embedded in the $D_4$ configuration
\be
& ds_{D_8}^2=\fft{2}{3(1+z^2)^{\fft{5}{6}}}\big(dz^2+(1+z^2)^{\fft{4}{3}}\eta_{\mu\nu}dx^\mu dx^\nu +\fft{9}{4}(1+z^2) d\Omega_4^2\big)\,,\quad z\in (-\infty,\infty)\,,\nn\\
& e^{\phi}=\sqrt{\frac{2}{3}} \sqrt[4]{z^2+1} g_s\,,\quad F_4=dC_3=\fft{2\pi N}{{\rm Vol}_4}\epsilon_4\,,
\ee
on which we have the DBI action\footnote{Our convention of gauge field is different from \cite{Sakai:2004cn,Sakai:2005yt}: $F^{\rm here}_{\mu\nu}=\partial_\mu A_\nu-\partial_\nu A_\mu-i [A_\mu,A_\nu]$, $F^{\rm there}_{\mu\nu}=\partial_\mu A_\nu-\partial_\nu A_\mu+ [A_\mu,A_\nu]$.}
\be
S=-\fft{1}{(2\pi)^8 \ell_s^9}\int d^9x\sqrt{-g}\,e^{-\phi} \Big(-{\rm Tr}\big\{{\rm det}\big[g_{MN}-2i\pi\alpha^\prime F_{MN}\big]\big\}\Big)^{\fft{1}{2}}\,.
\ee
Since we have the complete picture from the string theory, we can easily keep track of the power counting (where we keep the leading KK tower mass to be $M_{KK}=1$ for simplicity) \cite{Sakai:2004cn,Sakai:2005yt}
\be
\alpha^\prime=\ell_s^2=\fft{9}{2\lambda}\,,\quad g_s =\fft{\lambda^{\fft{3}{2}}}{3\sqrt{2}N\pi}\,,\quad \lambda=g_{{\rm YM},c}^2 N\,.
\ee
$\lambda$ is the 't Hooft coupling, where $g_{{\rm YM},c}$ is the Yang-Mills coupling for the colour sector on $D_4$ brane. The weakly-coupled supergravity regime is only valid for $\lambda\rightarrow\infty, N\rightarrow\infty$.

To map this model to our bottom-up EFT \eqref{eq: Lag YM}, we cut the brane in half for $z\in (-\infty,0)$ and introduce an additional gauge field to compensate for the contribution from the other half, where $z_{\rm UV}=-\infty$ and $z_{\rm IR}=0$. We find 
\be
\varphi=\frac{9 \sqrt{3} (1+z^2)^{\fft{1}{12}}}{4 \sqrt{2}}\,,\quad a(z)=\sqrt{\frac{2}{3}} (1+z^2)^{\fft{1}{4}}\,,\quad b(z)=\frac{\sqrt{\frac{2}{3}}}{\left(1+z^2\right)^{5/12}}\,,\quad g_{\rm YM}^2=\frac{486 \pi ^3}{\lambda  N}\,.
\ee
Besides, expanding in $\alpha^\prime$ yields \cite{Hagiwara:1981my}
\be
g_H=0\,,\quad \alpha_1=- \pi^2(\alpha^\prime)^2\,,\quad \alpha_2=4\pi^2 (\alpha^\prime)^2\,.
\ee
Using the dictionary \eqref{eq: dictionary}, we find
\be
f_\pi^2=\fft{\lambda N}{54\pi^4}\,,\quad l_1= -f_\pi^2\big(0.122985-\fft{4.62298}{\lambda^2}\big)\,,\quad l_2=0.122985 f_\pi^2\,.
\ee
At the leading order when $\lambda\rightarrow\infty$, we reproduce the results of \cite{Sakai:2004cn,Sakai:2005yt}. We can immediately see that the string correction causes the Wilson coefficients to deviate from the Skyrme model. To compare with the large-$N$ $\chi$PT bound, we should examine $\tilde{g}_{2}^\prime$ and $\tilde{g}_2$
\be
\tilde{g}_2=0.491942 \fft{M^2}{M_{KK}^2}+\fft{18.4919 M^2}{\lambda^2 M_{KK}^2}\,,\quad \tilde{g}_2^\prime=1.96777 \fft{M^2}{M_{KK}^2}\,.
\ee
At leading order, this is constrained to be below the kink for $M^2\leq 0.68 M_{KK}^2$. Indeed, the KK spectroscopy analysis suggests that the $\rho$ mass is $M_\rho^2=0.67 M_{KK}^2$! The string correction then pushes $\tilde{g}_{2}^\prime$ and $\tilde{g}_2$ upwards from the boundary, ensuring they still fall within the allowed region of Fig \ref{fig: g20g21plot}. We thus conclude that, even when including the leading string correction, we do not identify problems with the Witten-Sakai-Sugimoto model.

\subsubsection{Flat and AdS hard wall models}

Now we move to constrain two known holographic QCD models with $\varphi=1$. We focus on two models, one is constructed in the flat space as RS scenario \cite{Son:2003et}, another is the hard wall model \cite{Hirn:2005nr} constructed in AdS. They are both clearly explained in \cite{Hirn:2005nr}. We believe that our discussions can be generalized to other holographic QCD models (with possible modifications on how the chiral symmetry is breaking), like the soft wall models, where the dilaton $\varphi$ is nontrivially turned out \cite{Karch:2006pv,Ballon-Bayona:2020qpq,Ballon-Bayona:2021ibm,Afonin:2022qby}.

\begin{figure}[h]
\begin{subfigure}{.4\textwidth}
\centering
\includegraphics[height=0.32\textheight]{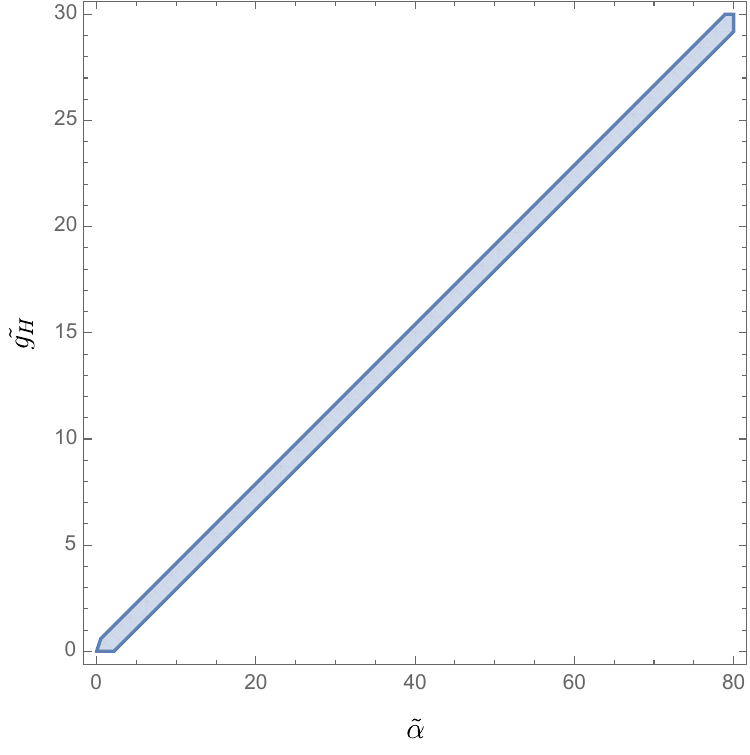} 
\caption{}
\label{subfig: flat}
\end{subfigure}
\hfill
\begin{subfigure}{.5\textwidth}
\centering
\includegraphics[height=0.32\textheight]{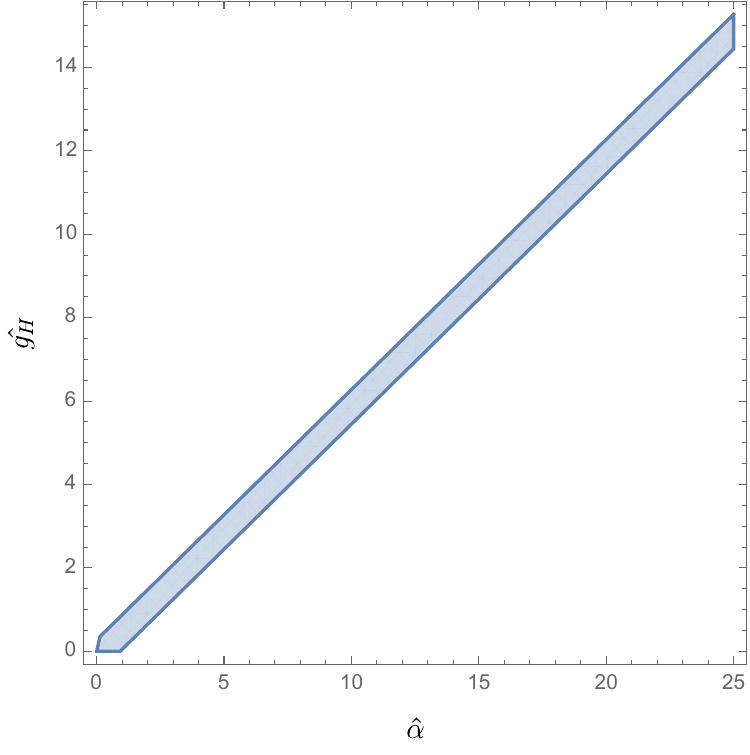}  
\caption{}
\label{subfig: AdS}
\end{subfigure}
\caption{(a) The bounds on holographic QCD supported by the model in flat space \cite{Son:2003et}  (b) The bounds on holographic QCD from hard wall model in AdS \cite{Hirn:2005nr}. All plots are still extending like a thin river.
}
\label{fig: holographic QCD bounds}
\end{figure}

\begin{itemize}

\item Flat space scenario \cite{Son:2003et}

For this model, we consider
\be
a(z)=b(z)=\varphi(z)=1\,,\quad z\in (0,z_{\rm IR}]\,,
\ee
where we fix the UV brane to be $z_{\rm UV}=0$, and $z_{\rm IR}\sim 1/\Lambda_{\rm QCD}$. We have
\be
& f_\pi^2=\fft{2}{g_{\rm YM}^2 z_{\rm IR}}\,,\quad l_1=\frac{15 \alpha +40 g_H z_{\text{IR}}^2-16 z_{\text{IR}}^4}{240 g_{\text{YM}}^2 z_{\text{IR}}^3}\,,\quad l_2=\frac{2 z_{\text{IR}}^2-5 g_H}{30 g_{\text{YM}}^2 z_{\text{IR}}}\,,\nn\\
& \tilde{g}_2=\frac{\pi ^2 \left(15 \alpha -40 g_H z_{\text{IR}}^2+16 z_{\text{IR}}^4\right)}{480 z_{\text{IR}}^4}\,,\quad \tilde{g}_2^\prime=\frac{1}{15} \pi ^2 \left(2-\frac{5 g_H}{z_{\text{IR}}^2}\right)\,,
\ee
where we have used $M^2\sim M_\rho^2=\pi^2/(4z_{\rm IR}^2)$ \cite{Son:2003et}. Besides, we denote $2\alpha_1+\alpha_2=\alpha$, which is the unique combination appears. We can easily observe some simple linear bounds
\be
  \tilde{g}_2^\prime >0 \rightarrow \tilde{g}_{H}=\fft{g_H }{z_{\rm IR}^2}\leq \fft{2}{5}\,,\quad \tilde{g}_2\geq \fft{1}{4} \tilde{g}_2^\prime\rightarrow \tilde{\alpha}=\fft{\alpha }{z_{\rm IR}^4} \geq 0\,.
\ee 
A more complete exclusion plot is depicted in Fig \ref{subfig: flat}, which looks like a thin river.
\end{itemize}

\begin{itemize}
\item Hard wall in AdS \cite{Hirn:2005nr}

For this model, we consider
\be
a(z)=b(z)=\fft{R_{\rm AdS}^2}{z^2}\,,\quad \varphi(z)=1\,,\quad z\in (0,z_{\rm IR}]\,.
\ee
We obtain
\be
& f_\pi^2 =\fft{4 R_{\rm AdS}}{g_{\rm YM}^2 z_{\rm IR}^2}\,,\quad l_1=\frac{24 \alpha +40 R_{\text{AdS}}^2 g_H-11 R_{\text{AdS}}^4}{192 R_{\text{AdS}}^3 g_{\text{YM}}^2}\,,\quad l_2=\frac{11 R_{\text{AdS}}^2-40 g_H}{192 R_{\text{AdS}} g_{\text{YM}}^2}\,,\nn\\
& \tilde{g}_2=-\frac{1.20483 g_H}{R_{\text{AdS}}^2}+\frac{0.722898 \alpha }{R_{\text{AdS}}^4}+0.331328\,,\quad \tilde{g}_2^\prime=1.32531\, -\frac{4.81932 g_H}{R_{\text{AdS}}^2}\,,
\ee
where we have used $M^2\sim M_\rho^2 \sim 5.78/z_{\rm IR}^2$ \cite{Hirn:2005nr}. Interestingly, in general we have two parameters $z_{\rm IR}$ and $R_{\rm AdS}$, but the resulting bounds suggest that $1/R_{\rm AdS}$ rather than $1/z_{\rm IR}$ is the cut-off for bulk EFT. We have similar simple bounds
\be
  \tilde{g}_2^\prime >0 \rightarrow \hat{g}_{ H}=\fft{g_H}{R_{\rm AdS}^2} \leq \fft{11}{40}\,,\quad \tilde{g}_2\geq \fft{1}{4} \tilde{g}_2^\prime\rightarrow \hat{\alpha}=\fft{\alpha}{R_{\rm AdS}^4} \geq 0\,.
\ee
The exclusion plot Fig \ref{subfig: AdS} also shows a thin river.

\end{itemize}

It is important to note that the dictionary provides bounds uniformly scaled by $z_{\rm IR}$ or $R_{\rm AdS}$, both of which are the IR scales in QCD. From naive dimensional analysis, we expect them to be bounded by the bulk``string" scale $\Lambda$. This either means that this method yields bounds that are too weak, or that the IR RS branes disrupt the naive dimensional analysis. On the other hand, we found that even though we identify $R_{\rm AdS}$ with $z_{\rm IR}$, we can't reproduce the flat-space scenario from the AdS one, which suggests a breakdown of the flat-space limit of causality bounds \cite{Caron-Huot:2021enk}. It would be interesting to explore all these points in the future for a better understanding.

\section{Summary}
\label{sec: summary}

We reviewed the EFT bootstrap, especially for the positivity scenario. We demonstrated that the EFT bootstrap is essentially an infinite-dimensional SDP, where the optimization Lagrangian can be formulated. We built the dual problem of the EFT bootstrap using the crossing symmetric dispersive sum rules, which embrace the crossing symmetry without the IR danger, and thus it serves as a better version of the improved sum rules. For the primal problem of the EFT bootstrap, we adapted the S-matrix primal ansatz and optimized the target Wilson coefficients.

We then applied the EFT bootstrap program to large-$N$ $\chi$PT, which is a low-energy pion EFT from the chiral symmetry breaking of large-$N$ QCD. Due to the large-$N$ limit, the positivity EFT bootstrap is sufficiently strong to carve out the allowed EFT space. Our dual bounds match with earlier literature \cite{Albert:2022oes,Fernandez:2022kzi}, and we demonstrated that the primal bounds are also converging to the dual rigorous bounds. This is consistent with the strong duality of SDP. We then focused on some converged bounds and used the primal solutions to extract the physical spectrum and S-matrix that saturate those bounds. By doing this, we confirmed some of the analytic rule-in amplitudes studied in \cite{Albert:2022oes,Fernandez:2022kzi}. Interestingly, for the Skyrme bound, we also observed a mysterious heavy Regge trajectory, which seems to suggest meta-stable exotic states with heavy quarks. In addition, we showed that the Regge behaviour of the primal ansatz does not affect the bounds, if it stays below the assumed Regge boundedness. This is consistent with SDP, as the dual problem is only sensitive to the Regge boundedness rather than the explicit Regge behaviour. Eventually, we incorporated the upper bound of the unitarity, i.e., the linear unitarity EFT bootstrap to confirm that the large-$N$ limit is consistent.

In the end, we considered the holographic QCD models, which are EFTs of gauge fields in $5D$ and correspond to large-$N$ QCD in $4D$. Typically, we included the higher derivative terms in the bulk and showed that they give rise to the general chiral Lagrangian up to $\mathcal{O}(p^4)$. We demonstrated that the Witten-Sakai-Sugimoto model with string corrections gives rise to a large-$N$ $\chi$PT within the allowed EFT region. Besides, for bottom-up models like the flat-space RS model and AdS hard-wall model, we translated the large-$N$ $\chi$PT to constrain the higher derivative couplings of ${\rm Tr}\, F^3$ and ${\rm Tr}\,F^4$ terms.

There are several aspects that deserve further investigations. From formal aspect, it would be interesting to build more precise relation between different methods for EFT and S-matrix bootstrap, include the SDP we reviewed, the moment problem \cite{Chiang:2021ziz,Chiang:2022jep,Chiang:2022ltp}, geometric function \cite{Raman:2021pkf}, iterative algorithm \cite{Tourkine:2023xtu,Tourkine:2021fqh} and machine learning approach \cite{Dersy:2023job}. Focusing on the SDP perspective, typically, we state that the crossing symmetric sum rules are the more natural tools to understand the loop effects on the EFT bounds, since they are free of forward-limit issues. It is then interesting to make this statement concrete by using the crossing symmetric sum rules to study the scalar EFT, $\chi$PT and other EFTs, with one and even two loop effects, trying to make the results of \cite{Bellazzini:2021oaj,Chala:2021wpj,Li:2022aby} sharp. For this exploration, it is also important to understand the primal-dual convergence when there are loops and the nonlinear unitarity is utilized. For example, we can apply the nonlinear unitarity bootstrap to real $\chi$PT like \cite{Guerrieri:2020bto,He:2023lyy}, and if the primal bounds and dual bounds converge, we can then probably extract the real QCD physics in the UV. Such analysis may also be extended to other important EFTs, like the standard model EFT \cite{Remmen:2019cyz,Bi:2019phv,Li:2022tcz,Ghosh:2022qqq,Chen:2023bhu}, gravitational EFT \cite{Caron-Huot:2022ugt,Caron-Huot:2022jli,Noumi:2022wwf,CarrilloGonzalez:2022fwg,Chiang:2022jep,deRham:2021bll}, QCD string EFT \cite{EliasMiro:2019kyf}, etc., helping us gain more information about their low-energy space as well as their possible UV completions. 

Particularly for large-$N$ $\chi$PT, it remains puzzling to us that the Skyrme model is problematic above the kink, since the Skyrme model is a good phenomenological model for understanding many aspects of nuclear physics, e.g., \cite{Klebanov:1985qi,Callan:1985hy}. It would be interesting to understand, microscopically, how the Skyrme model goes wrong above the kink. A possible route is to study the pion-nuclei scattering, which can be described as pion fluctuations around the Skyrmion \cite{Saito:1985wd,mattis1985baryon} (which are solitons of the Skyrme model and serve as the baryon \cite{adkins1983static,witten1979baryons}), and to detect if there are any causality violations like time advance \cite{Camanho:2014apa}. Besides, it is also interesting to understand where our constraints on holographic QCD models come from in the bulk. The constraints may again arise from classical causality, and techniques from \cite{deRham:2021bll,Chen:2021bvg,CarrilloGonzalez:2022fwg,CarrilloGonzalez:2023cbf,Chen:2023rar} would then be useful. This investigation may also be generalized to other RS scenarios, which provide the standard model EFTs.

\section*{Acknowledgements}	

We thank Simon Caron-Huot and Rohan Moola for their initial collaboration on the numerical exploration of the primal S-matrix bootstrap. We are also grateful to Jan Abert, Simon Caron-Huot, Miguel Correia, Julio Para-Martinez, David Simmons-Duffin,  Zhuo-Hui Wang and Shuang-Yong Zhou for insightful discussions, and to Jan Abert for sharing their raw data, and to Simon Caron-Huot for valuable comments on the draft. Additionally, we would like to thank Ofer Aharony, Gabriel Cuomo, Leonardo Rastelli, Victor Rodriguez, Igor Klebanov, Juan Maldacena, and Pedro Vieira for useful conversations. YZL is supported in parts by the Simons Foundation through the Simons Collaboration on the Nonperturbative Bootstrap and by the US National Science Foundation under Grant No. PHY- 2209997. The computations presented here were partly conducted using the Narval and Graham clusters supported by Calcul Qu\`ebec and Compute Canada, and partly conducted in the Resnick High Performance Computing Center, a facility supported by Resnick Sustainability Institute at the California Institute of Technology.

\pagebreak
	
\appendix

\section{EFT bootstrap as SDP: more}
\label{app: EFT}
In this appendix, we formulate the linear unitarity bootstrap and the nonlinear unitarity bootstrap as SDP.

\subsection{Linear unitarity}

For linear unitarity bootstrap, we write the Lagrangian as follows
\be
&\mathcal{L}_g= -\mathcal{F}\circ B(p^2)-2(1-S+i T)\int_{M^2}^{\infty} \sum_\rho Y_\rho^{(2)}(s)\,,\nn\\
& \mathcal{F}\circ B(p^2)=-g+\int_{M^2}^{\infty} ds \sum_\rho {\rm Im}\, a_\rho(s) \big(Y_\rho^{(1)}(s)-Y_\rho^{(2)}(s)\big)\,.
\ee
This SDP reads
\begin{itemize}
\item[] Primal
\be
\text{Minimize}\,\, g\,,\quad \text{Subject to}\,\,  0\leq {\rm Im}\,a_\rho(s) \leq 2\,, S_{\rm free}=1\,.
\ee
\item[] Dual
\be
\text{Maximize}\,\quad -2\int_{M^2}^{\infty} \sum_\rho Y_\rho^{(2)}(s)\,,\quad \text{Subject to}\,\, Y_\rho^{(i)}(s)\succeq0\,.
\ee
\end{itemize}

\subsection{Nonlinear unitarity}

The nonlinear unitarity bootstrap is more subtle. From primal side, we have further requirement for real part of $a_\rho(s)$, however, such object does not appear in our dispersive sum rule. Nevertheless, one can shift the dispersion relation to finite $|s_0|>M$, we then have, for example for spin-$2$ sum rule
\be
& B_2(s_0,p^2)=\oint \fft{ds}{4\pi i} \fft{2s+t}{(s_0-s)^2(s_0+s+t)^2}=0\rightarrow\nn\\
&   B_2(s_0,p^2)\Big|_{\text{low arc}} + \fft{\mathcal{M}(s_0,t)-\mathcal{M}(u_0,t)+(2s_0+t) \partial_{s_0}\mathcal{M}(s_0,t)}{2s_0+t}+ B_2(s_0,p^2)\Big|_{\rm high}=0\,.\label{eq: re ex1}
\ee
We can then invert $\mathcal{M}(s_0,t)$ to have $a_\rho(s)$, which is expressed in terms of complicated integral over low-energy contributions and UV part of the sum rules; schematically, we may have
\be
{\rm Re}\,a_\rho(s)\sim - i \,{\rm Im}\,a_\rho(s) + \int_{M^2}^{\infty} ds_1 \int ds_2 \mathcal{Y}_\rho(s_1,s_2) {\rm Im}\,a_\rho(s) + \text{low energy contribution}\,.\label{eq: re ex2}
\ee
This type of relation can be used to build the functionals acting on ${\rm Re}\,a_\rho(s)$ by functionals acting on ${\rm Im}\,a_\rho(s)$ with double integrals. We then can write down the Lagrangian for nonlinear unitarity
\be
& \mathcal{L}_g=-\mathcal{F}\circ B(p^2)-2(1-S+i T) \int_{M^2}^\infty \sum_\rho Y_\rho^{22}(s)-2\int_{M^2}^\infty ds \sum_\rho {\rm Re}\,a_\rho(s) Y_\rho^{12}(s)\,,\nn\\
&  \mathcal{F}\circ B(p^2)=-g+\int_{M^2}^{\infty} ds \sum_\rho {\rm Im}\, a_\rho(s) \big(Y_\rho^{11}(s)-Y_\rho^{22}(s)\big)\,.
\ee
In this Lagrangian, $Y^{11}$ and $Y^{22}$ can be constructed using the standard dispersive sum rules, while $Y_\rho^{12}(s)$ can only realized using intricate operations like \eqref{eq: re ex1} and \eqref{eq: re ex2} and contain double integral. 

This SDP then reads

\begin{itemize}
\item[] Primal:
\be
\text{Minimize}\,\, g\,,\quad \text{Subject to}\,\,  \mathcal{S}=
\left(
\begin{array}{cc}
 {\rm Im}\,a_\rho(s) & {\rm Re}\,a_\rho(s) \\
  {\rm Re}\,a_\rho(s) & 2- {\rm Im}\,a_\rho(s) \\
\end{array}
\right)\succeq 0\,.
\ee
\item[] Dual:
\be
\text{Maximize}\,\quad &-2\Big(\int_{M^2}^{\infty} \sum_\rho Y_\rho^{22}(s)+\int_{M^2}^{\infty} \sum_\rho {\rm Re}\,a_\rho(s) Y_\rho^{12}(s)\Big)\,,\quad \text{Subject to}\nn\\
& Y=\left(
\begin{array}{cc}
 Y_\rho^{11}(s) & Y_\rho^{12}(s) \\
 Y_\rho^{12}(s) & Y_\rho^{22}(s) \\
\end{array}
\right)\succeq 0\,.
\ee
\end{itemize}

Unfortunately, a more concrete dual example of this type of bootstrap is beyond the scope of this paper. We refer the readers to relevant discussions in \cite{Henriksson:2021ymi}. It would be interesting to explicitly realize the dual algorithm we propose here and compare with other dual algorithm in the future \cite{Henriksson:2021ymi}.

\section{Projectors of irreducible representation in SU$(N_f)$}
\label{app: projector}

In this appendix, we record all the projectors of irreducible representation in SU$(N_f)$ \cite{BandaGuzman:2020wrz} that we used to organize the pion amplitudes.

\be
& P^0_{abcd}=\fft{1}{2N_f}\delta_{ab}\delta_{cd}\,,\quad P^{{\rm adj}_{\rm A}}_{abcd}=-\fft{1}{N_f}f^{abe}f^{ecd}\,,\quad P^{{\rm adj}_{\rm S}}_{abcd}=\fft{N_f}{N_f^2-4}d_{abe}d_{ecd}\,,\nn\\
& P^{\bar{a}s\oplus\bar{s}a}_{abcd}=-\fft{1}{2}\big(\delta_{ac}\delta_{bd}-\delta_{ad}\delta_{bd}\big)-\fft{1}{N_f}f_{abe}f_{ecd}\,,\nn\\
& P^{\bar{s}s}_{abcd}=\fft{N_f+2}{4N_f}\big(\delta_{ac}\delta_{bd}+\delta_{ad}\delta_{bc}\big)-\fft{N_f+2}{2N_f(N_f+1)}\delta_{ab}\delta_{cd}-\fft{N_f+4}{4(N_f+2)}d_{abe}d_{ecd}+\fft{1}{4}\big(d_{abe}d_{ecd}+d_{cbe}d_{ead}\big)\,,\nn\\
& P^{\bar{a}a}_{abcd}=\fft{N_f-2}{4N_f}\big(\delta_{ac}\delta_{bd}+\delta_{ad}\delta_{bc}\big)+\fft{N_f-2}{2N_f(N_f-1)}\delta_{ab}\delta_{cd}+\fft{N_f-4}{4(N_f-2)}d_{abe}d_{ecd}-\fft{1}{4}\big(d_{abe}d_{ecd}+d_{cbe}d_{ead}\big)\,.
\ee

\section{Fixing-parameter method}
\label{app: fixing-parameter}

In this appendix, we explain the ``fixing-parameter'' method for nonlinearly bounding two Wilson coefficients, as a complementary of ``angle-searching'' method that was described in the main text.

The Lagrangian of ``fixing-parameter method'' is
\be
\mathcal{L}= g_1 + \lambda_1 (g_0-1)+\lambda_2 (g_2-g_2^\ast)-\int_{M^2}^{\infty}ds\sum_J {\rm Im}\,a_J (s) Y_J(s)\,.
\ee
The interpretation is that we fix $g_2=g_2^\ast$ and bound $g_1$ in the unit of $g_0$. From dual method, this corresponds to having
\be
\mathcal{F}\circ B(p^2)\Big|_{\rm low}=-g_1-\lambda_1 g_0-\lambda_2  g_2\,,\quad {\rm maximize}\,\, \lambda_1+\lambda_2 g_2^\ast\,,
\ee
which is precisely the fixing-parameter dual method used in \cite{Zhang:2021eeo}. On the primal side, however, the implementation is a bit subtle. $g_0=1$ is the normalization condition in the primal algorithm, then how should we address $g_2=g_2^\ast$? Recall that any Wilson coefficients are linear combinations of the coefficients in primal ansatz that we aim to solve, this indicates that $g_2=g_2^\ast$ put more constraints on the primal ansatz. Effectively, the primal ansatz is then degenerate and the coefficients there are no longer all independent. The strategy is to make an matrix $R$ to reduce the anstaz to independent subspace so that
\be
R\cdot g_2=g_2^\ast \, R\cdot g_0\,,
\ee
where we understand $g_2$ and $g_0$ as vectors spanned by the primal ansatz. Thus we can play with the effective Lagrangian
\be
\mathcal{L}=R\cdot g_1 + \lambda (R\cdot g_0-1)-\int_{M^2}^{\infty}ds\sum_J {\rm Im}\big(R\cdot a_J (s) \big) Y_J(s)\,.
\ee

\bibliographystyle{JHEP}
\bibliography{refs}

\end{document}